\documentclass[12pt,draftcls,onecolumn]{IEEEtran}

\usepackage{cite}

\usepackage[pdftex]{graphicx}

\usepackage[cmex10]{amsmath}
\usepackage{amssymb}

\usepackage{algorithm}
\usepackage{algorithmic}

\usepackage{array}

\usepackage[pdftex]{graphicx}
\usepackage{subfigure}

\usepackage{url}

\usepackage{multirow}

\usepackage{color}
\definecolor{red}{rgb}{1.0,0.0,0.0}
\definecolor{blue}{rgb}{0.0,0.0,1.0}

\usepackage{cases}

\usepackage[normalem]{ulem} 
\usepackage{soul,xcolor} 

\begin{document}

\setstcolor{red}

%
\title{Scalable RAN Virtualization in Multi-Tenant LTE-A Heterogeneous Networks}

\author{Georgia~Tseliou,~\IEEEmembership{Student Member,~IEEE,}
        Ferran~Adelantado,~\IEEEmembership{Member,~IEEE,}
        and~Christos~Verikoukis,~\IEEEmembership{Senior Member,~IEEE}
\IEEEcompsocitemizethanks{Copyright (c) 2015 IEEE. Personal use of this material is permitted. However, permission to use this material for any other purposes must be obtained from the IEEE by sending a request to pubs-permissions@ieee.org.}      
\thanks{G. Tseliou and F. Adelantado are with the Open University of Catalonia (UOC), Barcelona, Spain, e-mail: (\{gtseliou, ferranadelantado\}@uoc.edu).}
\thanks{C.Verikoukis is with the Telecommunications Technological Centre of Catalonia (CTTC), Castelldefels, Spain, e-mail: (cveri @cttc.es).}
}

\maketitle

\begin{abstract}
Cellular communications are evolving to facilitate the current and expected increasing needs of Quality of Service (QoS), high data rates and diversity of offered services. Towards this direction, Radio Access Network (RAN) virtualization aims at providing solutions of mapping virtual network elements onto radio resources of the existing physical network. This paper proposes the Resources nEgotiation for NEtwork Virtualization (RENEV) algorithm, suitable for application in Heterogeneous Networks (HetNets) in Long Term Evolution-Advanced (LTE-A) environments, consisting of a macro evolved NodeB (eNB) overlaid with small cells. By exploiting Radio Resource Management (RRM) principles, RENEV achieves slicing and on demand delivery of resources. Leveraging the multi-tenancy approach, radio resources are transferred in terms of physical radio Resource Blocks (RBs) among multiple heterogeneous base stations, interconnected via the X$2$ interface. The main target is to deal with traffic variations in geographical dimension. All signaling design considerations under the current Third Generation Partnership Project ($3$GPP) LTE-A architecture are also investigated. Analytical studies and simulation experiments are conducted to evaluate RENEV in terms of network's throughput as well as its additional signaling overhead. Moreover we show that RENEV can be applied independently on top of already proposed schemes for RAN virtualization to improve their performance. The results indicate that significant merits are achieved both from network's and users' perspective as well as that it is a scalable solution for different number of small cells.

\end{abstract}


\IEEEpeerreviewmaketitle

\section{Introduction}

Research conducted in the last years reveals that cellular networks will have to become heterogeneous and denser to meet the envisaged demands \cite{Ericsson2013}. Considering that operating infrastructure is a significant cost for operators, the densification of access networks and the necessity to reduce the costs will lead to cooperation between them and, to the sharing of resources, including infrastructure sharing itself. In this context, the provision of solutions enabling the creation of logically isolated network partitions over shared physical network infrastructure should allow multiple heterogeneous virtual networks to coexist simultaneously and support resource aggregation. This concept defines the principle of network virtualization \cite{Carapinha2009} and explains why Radio Access Network (RAN) virtualization emerges as a key aspect of the future cellular Long Term Evolution-Advanced (LTE-A) networks.

Today's cellular networks have relatively limited support for virtualization. Thus, although Third Generation Partnership Project ($3$GPP) standardizes necessary functionalities to enable several Core Network (CN) operators to share one RAN \cite{2013n}, neither a detailed implementation of radio resource customization among them nor mechanisms to exploit the network heterogeneity of the dense multi-tier architectures, defined in the latest release of LTE-A, are provided \cite{2013}. Therefore, the particular definition of algorithms implementing RAN virtualization for radio resources in a multi-operator sharing architecture still remains an open issue. In this point, we define \textit{RAN virtaulization} according to \cite{7054720}, as the way \textit{``in which physical radio resources can be abstracted and sliced into virtual cellular network resources holding certain corresponding functionalities, and shared by multiple parties through isolating each other"}. In turn, \textit{network sharing} is defined as the sharing configuration where \textit{``multiple CN operators have access to a common RAN" \cite{2013n}}.

The main challenges that should be addressed by RAN virtualization in LTE-A are i) the capacity limitation imposed by resource allocation, ii) the complete isolation between multiple coexisting services, and iii) the additional signaling overhead of each proposed solution. These challenges are even more complex to tackle in dense multi-tier scenarios, where Small Cells (SCs), are characterized by reduced coverage areas and therefore make the scenario more prone to geographical traffic non-uniformities \cite{6824255}. Traffic load and deployment are the foremost aspects of investigating the potential effectiveness of RAN virtualization. Although research solutions proposed so far have been mainly focused on the virtualization of resources in each Base Station\footnote{Throughout the rest of this manuscript, the term BS is used to describe either a macro eNB or a small cell (SC). The exact name of the BS is defined in all the particular cases that require the exact distinction among them.} (BS), there is still a gap in the literature for solutions that abstract the available resources to deliver them to multiple tenant BSs, considering geographical traffic variations that can occur in heterogeneous scenarios.

This work, taking into account the gaps in the current literature, is aimed to shed light on the limitations of the existing LTE-A RAN virtualization solutions by coping with dense multi-tier networks, to meet the requirements by the operators. Specifically, our contribution is twofold. Firstly, we extend and modify our previous proposal from \cite{Tseliou2014(tobepresented)}, the Resources nEgotiation for NEtwork Virtualization (RENEV) algorithm, for dynamic virtualization of radio resources spread in a two tier topology. Motivated by the geographical traffic variations, we propose a solution where baseband modules of distributed BSs, interconnected via the logical point-to-point X$2$ interface, cooperate to reallocate radio resources on a traffic need basis. Our proposal is based on the concept of physical resources transfer, defined as the possibility of reconfiguring the Orthogonal Frequency Division Multiple Access (OFDMA)-based medium access of two BSs, to allow a BS to use a set of subcarriers initially allocated to another BS. Resource customization to various tenants, i.e., BSs, is conducted after appropriate signaling exchange. RENEV is a virtualization solution that abstracts resources, by customizing them in isolation among different Requesting BSs. Secondly, we identify the basic limitations and signaling overhead caused to the current $3$GPP LTE-A architecture. In that sense, RENEV is harmonized and adapted to be compatible with LTE-A multi-operator network sharing configuration. Additionally, an insight on the analysis of the additional signaling overhead is given, since it is a key issue for virtualization, particularly as the network planning becomes denser.

The remainder of this paper is organized as follows. Section~\ref{Related Work} introduces the state of the art and our contribution. Section~\ref{Proposed Scheme} provides an overview of the architectural elements and functions of the scenario and then the proposed algorithm is described. The signaling design considerations associated to each phase of our proposal in current $3$GPP architecture are presented in Section~\ref{Feasibility}. In Section~\ref{Throughput Analysis} we introduce the analytical framework for network's throughput and in Section~\ref{signaling analysis} we calculate the theoretical signaling overhead introduced by RENEV. Both experimental and analytical results are illustrated to show the performance of RENEV in Section~\ref{Performance Evaluation}. Finally, conclusions are given in Section~\ref{Conclusion}.

\section{State of the Art and Contribution} \label{Related Work}

Cellular network sharing among operators, is a key building block for virtualizing future mobile carrier networks. 3GPP has recognized the importance of supporting network sharing among operators by defining a set of architectural elements \cite{TS22951} and technical specifications \cite{2013n}. Two possible architectural network sharing configurations have been specified: the Gateway CN (GWCN) and the Multi-operator CN (MOCN). In GWCN configuration, CN operators share control nodes in addition to RAN elements whereas in MOCN, multiple control nodes owned by different operators are connected to a shared RAN. Throughout this manuscript, the infrastructure owner provides the underlying physical network whereas by referring to network operator, we denote every operator having its users connected to the RAN (without necessarily owning infrastructure). In both configurations the network sharing agreement between operators is transparent to the end users. Although, operators may share network elements (i.e., RAN/control nodes), radio resources virtualization is required to cover their actual requirements, in isolation per BS. Therefore, in both MOCN and GWCN sharing configurations, virtualization of resources is necessary in order to allow operators' users to have access to the complete set of available resources. Existing network virtualization techniques, can be grouped into solutions for the Evolved Packet Core (EPC) Network and the RAN \cite{6887287}. This paper is focused on the RAN of an heterogeneous LTE-A deployment, which, in turn, can be divided in \textit{dynamic resources' slicing} and \textit{spectrum sharing}. 

With regard to the dynamic resources' slicing, interesting proposals are presented in \cite{Kokku2013,Li2012,Jin2013}. CellSlice framework is proposed in \cite{Kokku2013} to achieve active RAN sharing by remotely controlling scheduling decisions without modifying BS's schedulers. Instead, in \cite{Li2012,Jin2013,6362137}, the authors present software defined cellular network architectures, allowing remote gateway level controller applications to perform resource slicing without modifying the BSs' Medium Access Control (MAC) schedulers. Such solutions express real-time, fine-grained policies based on subscribers attributes rather than network addresses and locations.

As for spectrum sharing \cite{4WARD,Panchal2013,Kokku2012,NEC2013,Costa-Perez2013,Guo2013}, the proposals are designed to adapt the radio interface of the eNB to traffic load variations of distinct virtual networks. This objective is achieved by allowing multiple virtual networks to share the spectrum allocated to a particular physical eNB. A preliminary approach for virtualizing a BS in LTE is described in \cite{4WARD}. A controlling entity called hypervisor was proposed in order to make use of apriori knowledge (e.g., user channel conditions, operator sharing contracts, traffic load etc.) to schedule the Resource Blocks (RBs) of a BS among different mobile operators. In addition, the authors of \cite{Panchal2013} evaluate several sharing options, ranging from simple approaches feasible in traditional infrastructure to complex methods requiring a specialized one. In advancing the basic BS virtualization, works \cite{Kokku2012,NEC2013} and \cite{Costa-Perez2013} introduce the concept of Network Virtualization Substrate (NVS) that operates closely to the MAC scheduler. NVS adopts a two-step scheduling process, one managed by the infrastructure provider for controlling the resource allocation towards each virtual instance of an eNB and the second controlled by each virtual instance itself providing scheduling customization within the allocated resources. Additionally, \cite{Guo2013} extends NVS solution by investigating the provision of active LTE RAN sharing with Partial Resource Reservation (PRR). In this scheme, each slice is guaranteed a specific minimum share of radio resources to be available to the operator that owns them. The remaining common part is shared among traffic flows belonging to different operators. 

Based on the presented state of the art, virtualization solutions proposed so far have been mainly focused on allocating resources, per operator/service, within a specific BS (\cite{Kokku2013,Kokku2012}). In particular, whereas in some proposals resources are dynamically sliced between services with different QoS characteristics (\cite{4WARD,Panchal2013,NEC2013}), in other proposals the same resources are virtualized and distributed among different operators with shared access to the same BS (\cite{Costa-Perez2013,Guo2013}). Such proposals are effective virtualization solutions to address the traffic dynamics in two aspects: service and operator dimensions. In the first case, the variety of services poses challenges to resource allocation, whereas the second dimension is really interesting since the distribution of traffic between different operators is not necessarily uniform. However, none of the aforementioned proposals is able to cope with dynamics in a third aspect of traffic: the geographical dimension.

Heterogeneous networks (HetNets) are characterized by dense deployment of BSs with different transmission power and overlapped coverage areas. In these scenarios, the densification of the network with low-power BSs (i.e., SCs) has clear impacts on the traffic load: i) the distribution of the traffic between BSs is not uniform  \cite{6824255,6692277}, and ii) the variability of traffic in the short-term, particularly in SCs, is high. As a consequence the overall capacity of the system is usually compromised by spatial non-uniformities. Therefore, even appropriate deployments, which are static in nature, are unable to optimally tackle the spatial variations of the traffic.

Accordingly, an efficient use of the available radio resources can be achieved if a proper coordination / negotiation of resources is carried out among the BSs. In \cite{Tseliou2014(tobepresented)}, we introduced a first approach of RENEV and applied it in a deployment consisting only of SCs (i.e., Home Evolved Universal Terrestrial Radio Access (E-UTRA) NodeBs (HeNBs)). In such environments, RENEV is responsible for reallocating / transferring radio resources by reconfiguring the OFDMA based radio interface in a decentralized manner. The innovation in \cite{Tseliou2014(tobepresented)} lies in the fact that the baseband part of the BSs is shared and a common Radio Resource Control (RRC) layer for a specific group of BSs is created in a coordinated way. RENEV is essentially designed to reconfigure the radio resources of two BSs in order to adapt the allocation of resources to the traffic dynamics of an operator. Thus, when there is a tenant BS without enough resources to serve the offered traffic, RENEV should find out if there are unused resources in other neighboring BSs, check if the unused resources could be reallocated, and finally reconfigure the medium access of the two BSs to reallocate them from one to the other (hereinafter also known as transfer of resources). In this scenario, the hierarchical or non-hierarchical operation of the nodes arises as a key aspect.

RENEV offers a complementary solution to the state of the art and covers gaps found therein by introducing a new dimension in RAN virtualization. Accordingly, we extend the proposal in \cite{Tseliou2014(tobepresented)}, by modifying it, allowing BSs that belong to two tiers to reallocate underutilized spectrum to other BSs. Our main contributions can be summarized as follows:

\begin{itemize}[
    \setlength{\IEEElabelindent}{\dimexpr-\labelwidth-\labelsep}
    \setlength{\itemindent}{\dimexpr\labelwidth+\labelsep}
    \setlength{\listparindent}{\parindent}
]
\item We introduce RENEV as a solution, that can be employed on former RAN virtualization proposals (e.g., NVS \cite{Costa-Perez2013} and PRR \cite{Guo2013}), in HetNet scenarios composed of two tiers, each one operating on different sets of subcarriers. In these scenarios the geographical traffic non-uniformities render the initial allocation of resources into the BSs insufficient; some BSs are more loaded than others, resulting into areas that require more resources. On the one hand, RENEV is a virtualization solution that customizes resource slices from a BS to another, based on the traffic requirements created by the participating operators. On the other hand, virtualization solutions proposed so far in the literature (e.g., NVS \cite{Costa-Perez2013} and PRR \cite{Guo2013}) only allow resources customization among operators/services within the same physical BS.

\item We demonstrate that RENEV could be applied independently on top of existing virtualization solutions (e.g., NVS \cite{Costa-Perez2013} and PRR \cite{Guo2013}), thereby guaranteeing its operation in multi-service multi-operator scenarios \cite{Panchal2013}. The implementation of RENEV does not impose additional constrains to the virtualization of resources within each tenant BS, proposed by the aforementioned solutions.

\item We analytically derive the upper bounds of the throughput with and without RENEV.

\item We provide the description and analytical model of the signaling introduced by RENEV, a key point in the dimension of the physical connections that support the logical X2 interface 
\end{itemize}

\section{Resources Negotiation for Network Virtualization (RENEV)} \label{Proposed Scheme}

\subsection{Scenario under consideration} \label{IIIA}

In this subsection, we introduce (i) the specific network sharing configuration where the resources virtualization by RENEV is applied and (ii) its architectural elements (i.e, the RAN nodes (BSs), the control nodes and their interconnecting interfaces). In our scenario different CN operators may connect to a shared RAN \cite{2013n}. We study the GWCN configuration, where different operators may also share the same control node. This sharing configuration, consists of a set of resources belonging to the RAN elements and need to be customized in isolation among users of multiple operators. Regarding the RAN elements (whose resources need to be virtualized), the underlying considered network is a residential region composed of an eNB and a number of open access mode SCs placed throughout its coverage area in clusters, close to each other, in random positions \cite{6056684}. The two tiers are initially assigned disjoint frequency bands \cite{2013i}; however by exploiting the concept of Carrier Aggregation (CA), both tiers can operate on the whole bandwidth [5]. Most RAN nodes maintain standardized connections to each other, for example, BSs are connected to their neighbors using the point-to-point, logical X$2$ interface to support a direct control and data information exchange. Furthermore, we focus on the downlink, where the RB is the basic time-frequency resources unit. In principle, any RB can be assigned to one or several BSs subject to interference limitations. The eNB is assumed to transmit with a fixed power per RB. The downlink transmitted power per RB is also fixed and equal among the SCs \cite{Ali2011}.

The BSs are connected to the EPC directly with the Mobility Management Entity (MME) or through an intermediate node, named Home eNB Gateway (HeNB GW) using the S$1$ interface \cite{2012}. These nodes manage BSs to provide a radio network. According to GWCN network sharing configuration \cite{2013n}, these control nodes are shared by different operators as defined by their Service Level Agreement (SLA). Therefore, this sharing configuration may host a scalable number of CN operators owning both CN and RAN nodes.

Based on \cite{2013}, three ways of interconnection of the tenant BSs arise: (i) a cluster of SCs (i.e., in our test case HeNBs) connected to the same HeNB GW, (ii) a group of eNBs connected to the same MME and (iii) a group of eNBs as well as SCs associated to the same MME. In the first and second case, the HeNB GW and the MME concentrate the control plane of the SCs and the eNBs respectively. In the last case the MME integrates the control plane of both types of BSs within a certain geographical area. Despite the different cases presented in \cite{2013}, from a BS's perspective all cases are identical in terms of signaling. This means that the message exchange from the BS-BS communication required by RENEV, is independent from the coverage area and the transmission power of a BS. Therefore, we consider equivalent the cases of message exchange between eNB-SC and SC-SC that is required when executing RENEV. Under these circumstances, in a scenario like this, we further assume that the involved BSs are necessarily deployed over the same geographical area, and therefore connected to the same control node (i.e., MME) which is shared by multiple operators.

\subsection{Radio Resource Management Functions}

The management of spectrum resources allocated to the BSs, relies on their control plane.

The control plane of a BS in LTE-A is logically divided in two entities: baseband and network modules, as defined in the standard in \cite{2013}. The former is responsible for bearer setup, to register users from each operator to the network via RRC protocol, whereas the latter connects the BS with the EPC. Radio Resource Management (RRM) is implemented in baseband module of a BS with primary goal to control the use of radio resources in the system, by ensuring QoS requirements of the individual radio bearers and minimization of the overall use of resources.

Focusing on the baseband module, two fundamental functions of the RRM jointly manage the resources of a BS: the Radio Bearer Control (RBC) and Radio Admission Control (RAC) \cite{2013}. On the one hand, RBC is responsible for the establishment, maintenance and release of radio bearers. When setting up a radio bearer, RBC considers the overall resource situation and QoS requirements of in-progress sessions \cite{2013h}. Correspondingly, it is involved in the release of radio resources at session termination. On the other hand, the task of RAC is to admit or reject the establishment requests for new radio bearers. RAC ensures high radio resource utilization by accepting bearer requests from operators as long as radio resources are available. At the same time, it ensures proper QoS for in-progress sessions by rejecting radio bearer requests when they cannot be accommodated \cite{2013a}. A new bearer will be built only if radio resource in the cell is able to maintain the QoS of the current sessions. It will be released at the end of the communication. Based on the role played by RBC and RAC, any RRM technique aimed to improve the efficiency in the dynamic allocation of the radio resources among BSs must interact with these two functions.

\subsection{Proposed Algorithm: RENEV} \label{Nomenclature RENEV}

In the scenarios described in Section \ref{IIIA}, traffic non-uniformities among BSs make resource allocation a challenging task. A dynamic coordination of radio resources is required to address such kind of variations. This is the objective of RENEV in these environments; customizing resources in terms of RBs, to satisfy new incoming user requests by multiple operators in tenant BSs, while supporting isolation among the reallocated resource slices.

Let us define the number of RBs initially allocated to a particular BS as $RB$, and the number of RBs required to serve the demand of its associated users as $u$. By definition, the number of available RBs in this specific BS, denoted as $r$, can be expressed as $r=RB-u$. As long as $r>0$, the tenant BS will be able to serve the offered traffic. Conversely, when $r<0$, the BS will start to degrade users' performance and block UEs' incoming attachment requests.

\begin{figure*}[htbp]
	\centering
	\includegraphics[width=\textwidth]{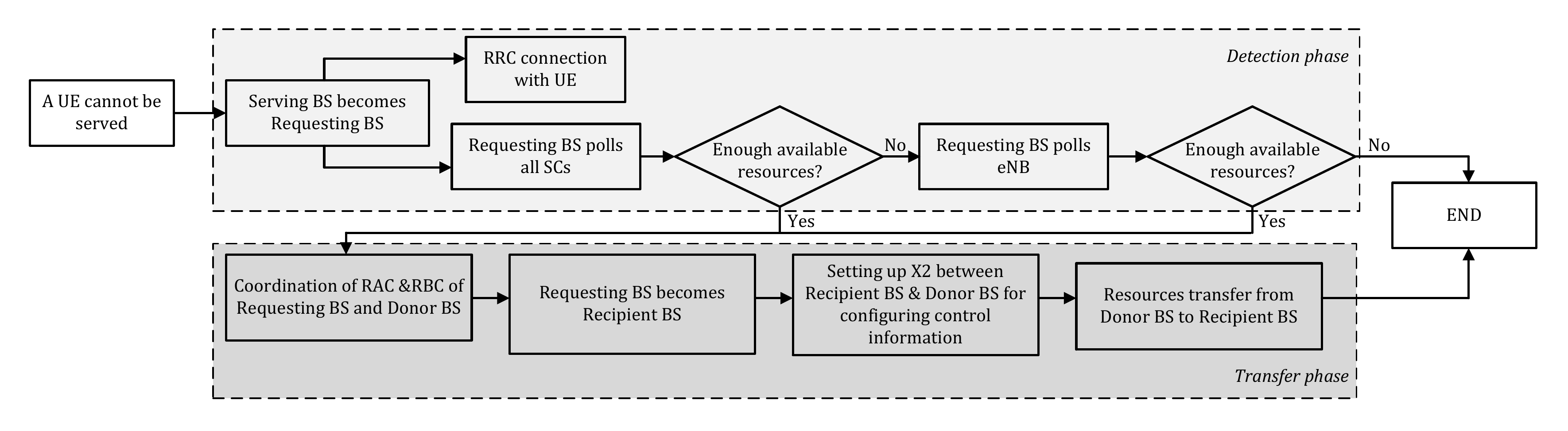}
	\caption{RENEV for a Heterogeneous Deployment.}
	\label{flowchart}
\end{figure*}

It is particularly worth noting that in HetNets the significant variability of the traffic among neighbouring BSs can lead to the paradox of having some BSs with $r<0$ and, at the same time, some other BSs with $r \gg 0$. RENEV is defined as the decentralized procedure intended to match the tenant BSs with $r<0$ and the ones with $r>0$, and manage the exchange of control messages to reconfigure the allocation of resources among them. For this reason, RENEV is divided into two sequential phases, as shown in Fig.~\ref{flowchart}.

First, the detection phase, where a BS with $r<0$ seeks among the neighbouring BSs if any of them has $r>0$. This search is carried out by polling one by one the neighbouring tenant BSs to figure out the amount of available resources. Subsequently, the transfer phase is only executed if the tenant BS with $r<0$ finds neighbouring BSs with $r>0$. This phase consists in re-configuring the two involved BSs according to the operators' traffic requirements. The details of each phase are stated below, and a proposal of the messages exchanged during the two phases is described in Section \ref{Feasibility}. Before proceeding with the details, we describe the basic nomenclature:
\begin{itemize}[
    \setlength{\IEEElabelindent}{\dimexpr-\labelwidth-\labelsep}
    \setlength{\itemindent}{\dimexpr\labelwidth+\labelsep}
    \setlength{\listparindent}{\parindent}
]
	\item \textbf{Serving BS:} is the node that a User Equipment (UE) is associated to and it is responsible for serving it.
	\item \textbf{Requesting BS:} is the node that, after receiving an access request from a UE, determines that the request cannot be accommodated with the available resources. It is precisely at this time, that the node takes the role of Requesting BS and triggers a requesting process among the neighboring BSs to figure out if there are unused resources.
	\item \textbf{Requested BS:} is the node that, after a neighboring Requesting BS triggers a requesting process, receives a request to inform about its unused resources.        
	\item \textbf{Donor BS:} is the node that, upon the completion of a requesting process triggered by a Requesting BS, is selected to transfer resources to this Requesting BS.
	\item \textbf{Recipient BS:} is the role taken by a Requesting BS after reconfiguring the radio interface to use the resources transferred from a Donor BS.
\end{itemize}

Since, in general, spectral efficiency of SCs is higher than spectral efficiency of eNBs, SCs play the role of Requesting BSs. SCs are usually needed for dense deployments in high-traffic environments and therefore, they are more prone to lack resources. This is the main difference in the scale of macrocells and SCs. Thus, if the eNB could play the role of Requesting BS, the RBs transferred from a SC to the eNB could not be reused by any other SC, resulting in a reduction of the capacity. In RENEV, RBs transferred by the eNB can be reused in more than one SC in the SCs tier, given that the involved SCs do not have overlapped coverage areas. When the imbalance between the demanded and the allocated resources comes to an end (i.e., the additional resources transferred by RENEV to a Requesting BS are no longer needed), the resources given by the Donor BSs (resulting from the execution of RENEV) reverts to the initial allocation. As a consequence, the role of Requested BS can be held either by SCs or an eNB. In that sense, SCs can be both Donor and Recipient BSs, whereas eNB is always a Donor BS.

\subsubsection{Detection phase} \label{III C 1}
If a user from an operator can be served by resources owned by the Serving BS (i.e., $r>0$), then it is served \cite{2013a}. Otherwise, the Serving BS, after setting up a RRC connection on the air interface with the user requiring service provision, triggers RENEV by adopting the role of Requesting BS. At this point the detection phase starts (see Fig.~\ref{flowchart}). Next, the Requesting BS scans the local network\footnote{The local network of a BS is defined as the set of BSs deployed in its vicinity. Generally, this local network consists of an eNB and a finite number of SCs under its coverage area.} to find a potential Donor BS by polling BSs around it. The polling procedure undertaken by the Requesting BS may itself be divided into two steps. First, the Requesting BS polls each neighbouring SC, one by one, to monitor the resources status of the SCs tier. Secondly, if there are not available resources in the SCs tier, the Requesting BS polls the macro eNB. After completing the requesting process, the Donor BS is selected among the set of Requested BSs according to two criteria: load and proximity.

\begin{enumerate}[
    \setlength{\IEEElabelindent}{\dimexpr-\labelwidth-\labelsep}
    \setlength{\itemindent}{\dimexpr\labelwidth+\labelsep}
    \setlength{\listparindent}{\parindent}
]
\item \textbf{Load:} The Requested BS with more unused resources is selected as the Donor BS. Yet, in order for the Donor BS to be able to accommodate possible further increase of the traffic demand in the short/mid-term future, a Requested BS can only become a Donor BS if the amount of remaining resources after the transfer is above a minimum threshold.
\item \textbf{Proximity:} For a set of Requested BSs likely to become the Donor BS, and if more than a single Requested BS has the same amount of unused resources, the Donor BS will be the BS with the minimum distance to the Requesting/Recipient BS. This criterion guarantees that the effect of the algorithm is geographically restricted to limit undesirable instability problems caused by the nature of the wireless medium.
\end{enumerate}

Regarding the implementation details of this phase, when a user is attached to the Requesting BS, the RRC connection establishment is used to make the transition from RRC Idle to RRC Connected mode. This transition is carried out before transferring any application data, or completing any signaling procedures, as shown in Fig. \ref{flowchart}. RRC establishment procedure is always initiated by the user but it can be triggered by the user or the network \cite{2013a}. When the Requesting BS scans the network to find a Donor BS, a coordinated control connection of their baseband parts is created via the X$2$ interface. Every time that a polling procedure between a Requesting BS and a Requested BS is carried out, two messages are exchanged through X$2$ interface (one from the Requesting BS to the Requested BS, and another one vice versa).

\subsubsection{Transfer phase} \label{III C 2}

Upon detecting the Donor BS, the transfer of resources from the Donor BS to the Recipient BS takes place via X$2$ interface. It is worth noting, that the exchange of BS configuration data over the link must be preceded by resetting the link resolving security issues.

In the proposed scheme, RAC and RBC functions, belonging to RRC layer of distinct neighbouring BSs, cooperate to provide seamless service to the end users (first action of the transfer phase, dark shaded in Fig.~\ref{flowchart}). We leverage the logical split of a BS into baseband and network modules and create a common RRC process among the Recipient BS and the Donor BSs. When the Requesting BS finds the Donor BS, RRC functions of the two nodes are enabled; RAC is responsible for checking if the node has available resources and RBC for establishing the radio bearer; it is in that moment that the Requesting BS becomes the Recipient BS. Next the medium access of two involved BSs is reconfigured and spectrum is lent by the Donor BSs through the control communication of the nodes. This process is seamless to end users since RRC connection is maintained with the initial Requesting BS and it is done without the participation of additional BSs or gateways. Finally, the Donor BS leases the demanded resources, which are used by the Recipient BS.

\subsection{Discussion on RENEV}

\subsubsection{RAN Virtualization Properties }

In this subsection, we introduce the key virtulization properties of RENEV, its main differences with conventional joint resource allocation solutions and how it can interact with already proposed RAN virtualization schemes. RENEV provides the virtualization features defined by 3GPP SA1 RSE requirements \cite{Costa-Perez2013} :
\begin{itemize}[
    \setlength{\IEEElabelindent}{\dimexpr-\labelwidth-\labelsep}
    \setlength{\itemindent}{\dimexpr\labelwidth+\labelsep}
    \setlength{\listparindent}{\parindent}
]
\item \textbf{Abstraction:} RENEV abstracts the radio resources belonging to a deployment into a pool; these resources are delivered on demand to each Requesting BS according to the operators' needs. In particular, abstraction of resources is accomplished by the communication between Requesting BS and Requested BS (as defined in Section~\ref{III C 1}). Instead of having the view of the physical radio resources (i.e., RBs) in each BS, RENEV after being triggered creates a set of virtual resources. This set consists of physical resources coming from different Donor BSs and it is accessible by various Requesting BSs according to the existing traffic non-uniformities.

\item \textbf{Isolation:} RENEV ensures a reserved portion of resources to each Requesting BS that triggers it, to meet the requirements of the operators, in this specific BS. Traffic, mobility and fluctuations in channel conditions of one Requesting BS do not affect the reserved resource allocations of other Requesting BSs. More specifically, isolation is achieved during the Transfer phase (defined in Section~\ref{III C 2}) where RENEV creates a logical common RRC process among the Recipient BS and the Donor BS. RAC and RBC functions for the Requested BS are enabled,  and the lent resources are seamlessly reserved to be used by a particular Recipient BS.

\item \textbf{Customization:} RENEV offers the flexibility to different operators having access to the sharing configuration, to conquer different part of the shared resources according to the actual requirements. Resource customization is attained during the Detection phase of RENEV (defined in Section~\ref{III C 1}). When a BS runs out of resources, RENEV is triggered so as resources can be allocated to the Requesting BS that needs them according to the specific traffic load conditions.

\item \textbf{Resource Utilization:} RENEV guarantees the efficient use of physical radio resources with a rational signalling burden for applying the solution onto the network. The medium access of each pair Requesting - Requested BSs is reconfigured during RENEV. Thus, the spare spectrum is lent by Donor BS through the control communication of the nodes.
\end{itemize}

\subsubsection{Differences with Joint Resource Allocation and Generic Resource Sharing}

The aforementioned properties distinguish in general virtualization solutions from conventional joint resource allocation ones. As defined in \cite{6797977}, the latter \textit{``apply a joint optimization approach (power control, channel allocation, and user association) for resource allocation in a multicell network, which can be invoked at the network planning stage or when the resource status changes"}. Although in such kind of solutions, resources are allocated among cells, the isolation property does not hold. Traffic, mobility and fluctuations in channel conditions of users of one entity affect the resources that would be given to other entities. In RENEV, the customization of resources among tenant Requesting BSs is performed on demand, with the target to serve as many users as possible, belonging to distinct network operators that share the RAN. In our solution, dedicated resources are served and locked per Requesting BSs to be allocated to a user of a certain participating operator. However, a conventional multi-cell joint resource allocation solution, does not isolate any resources for specific operators within the topology. Therefore in RENEV isolated slices of RAN can be assigned to Requesting BSs, to serve the traffic needs of users belonging to distinct operators.

It is important to differentiate RENEV from generic resource sharing approaches. That is because generic resource sharing among multiple operators can be performed with or without virtualization. To highlight the difference, let us consider the Spectrum Sharing (SS) scheme presented in [15]. It represents a traditional resource sharing approach that is done via a ``request and release of spectrum" method where the portions that can be allocated are fixed and it is performed into BS level. According to SS, a supply sector belonging to an operator allows access of a portion of its own carrier to a heavily loaded demand sector (i.e., leased sector) of another collocated operator. Unlike resource sharing via virtualization, this sharing procedure requires spectrum division and reconfiguration - during this process the operators' users are put in a suspended state. While this is a conventional case of resource sharing in a BS, when adding virtualization, the allocation of resources takes place dynamically and the reconfiguration process is not necessary because the supply and the leased sectors share a number of physical RBs. For example in the NVS \cite{Costa-Perez2013} case, this is because the BS scheduler is modified: this virtualization solution does not require the same operators to be collocated in order to share the resources, neither the suspended state for the users. The trade-off is the added complexity due to the BS MAC scheduler modification. Similarly, RENEV is also performing a virtualization of resources, but in a higher level. Altogether, RENEV achieves resource sharing among operators via virtulization in a process that does not take place in each specific BS but in a set of resources owned by a geographically constrained set of BSs.

\subsubsection{Interaction with existing Virtualization Proposals}

Also it is worth emphasizing that RENEV operates independently, on top of existing virtualization solutions, such as NVS \cite{Costa-Perez2013} and PRR \cite{Guo2013}. In general, each virtualization mechanism abstracts physical resources to a number of virtual resources, which are then delivered in isolation to different tenants. However, resource virtualization may appear in different levels and distinct solutions can exist that determine how resources are distributed: \textit{within each BS} and \textit{above the BS}.

NVS and PRR are indicative examples for virtualization within each BS. For instance if NVS is implemented, a particular BS will lack resources as soon as the traffic from an operator consumes all resources devoted to it \cite{Costa-Perez2013}; if PRR is applied, all the traffic load from an operator will be served as long as the shared part of resources belonging to particular BS, is nonempty \cite{Guo2013}. Thus, the needs or surpluses of resources within the BS vary, based on the aggregate traffic demand by the operators in a particular BS and how the resources are distributed within it. However looking at the top-down approach, RENEV also virtualizes resources, but from a higher network perspective (i.e., above the BS): instead of performing its tasks per BS level, RENEV abstracts and slices the physical RBs according to the spatial traffic non-uniformities. The delivery of these resources is done on demand according to these requirements as described in section \ref{Nomenclature RENEV}.

All in all, there is no need for explicit communication between RENEV and these solutions. RENEV can first be implemented to virtualize resources among Requesting BSs of the whole deployment based on the aggregate operators' requirements (due to geographical non-uniformities of their traffic). Then NVS \cite{Costa-Perez2013} and PRR \cite{Guo2013} may be applied in each particular BS, to customize the available resources (made accessible by RENEV when required) among operators.

\section{Signaling Design Considerations} \label{Feasibility}

The additional signaling overhead introduced in the network is a key aspect of the proposed solution, since it could limit its feasibility. This section is intended to analyze in detail the signaling messages exchanged in the network to implement RENEV, as well as its compliance with the current standards and architecture of LTE-A. A short discussion about the time scale of RENEV is also introduced.

Any procedure concerning the accommodation of a new user in a cell, starts with its attachment as explicitly defined in the standard \cite{2013a}. The attachment of a user to a new cell is characterized by two main processes: firstly, the communication between UE and Serving BS over air (i.e., Uu) interface, and secondly, the communication between Serving BS and the MME to exchange initial UE context setup over S$1$ interface. 

Regarding the message exchange over Uu interface, the user sends the attach request message to the Serving BS, as also defined in the standard \cite{2013a}. After sending the first message of random access procedure to the network, denoted as RACH preamble, RRC connection is established. The initial UE context setup, consists of an exchange of messages with the purpose of transferring UE context information from the MME to the Serving BS. These messages are exchanged over S$1$-AP application layer using SCTP. When the appropriate RRC transport container is received by the Serving BS, the establishment of a dedicated SCTP control stream on S$1$-MME is triggered \cite{2013m}. The described procedure is nonetheless subject to the availability of resources in the Serving BS. In that sense, RENEV aims to transfer resources from one BS to another to minimize the number of unsuccessful procedures. Therefore, RENEV should be executed after the UE attachment request and before the UE context exchange.

\begin{figure*}[htbp]
	\centering
	\includegraphics[width=\textwidth]{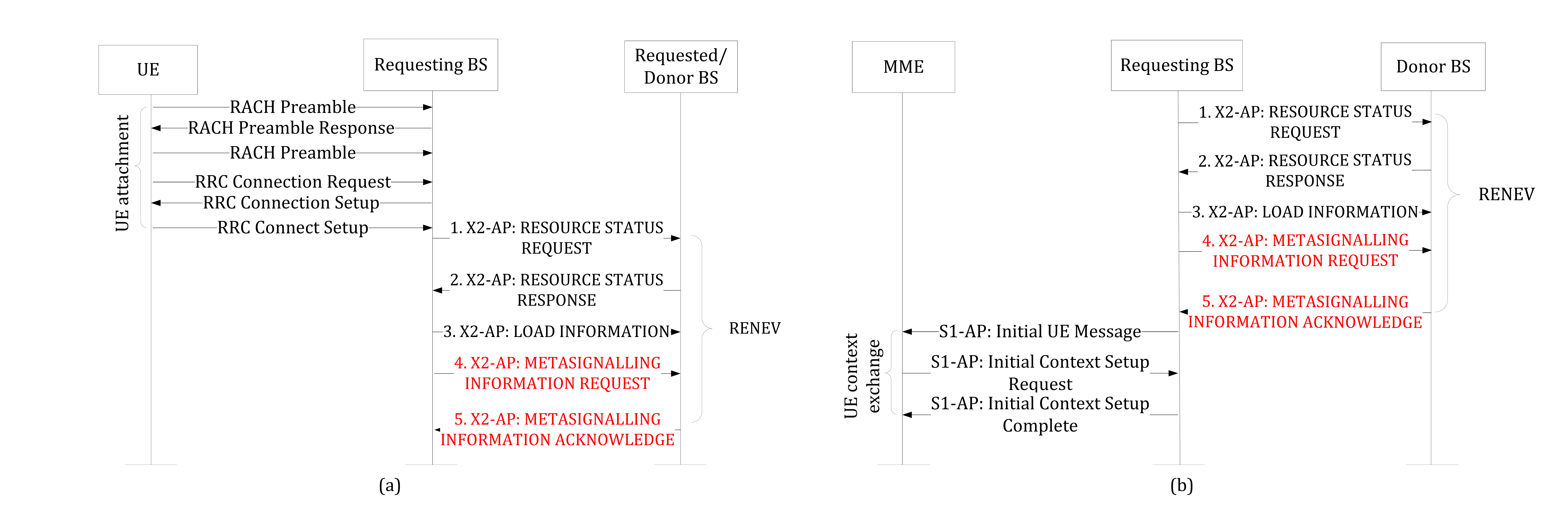}
	\caption{Call Flow of the messages for (a) UE Attachment and RENEV and (b) UE Context Exchange and RENEV.}	
	\label{fig_metasignalling}
\end{figure*} 

The direct communication between two BSs is conducted via X2, using the X$2$ Application Protocol (X$2$-AP) \cite{2013}. X$2$-AP messages are characterized by communication context identifiers and some specific parameters called Information Elements (IEs). These define the source and target BS, as well as characteristics of the transferred message. The messages required to implement RENEV are detailed below.

\subsection{Detection phase signaling}

When applying RENEV, the first process to carry out includes the polling procedure to detect spare resources (see Fig.~\ref{fig_metasignalling}(a), messages $1$, $2$ and $3$). During this operation, the Requesting BS scans the network to find the Donor BS, as shown in Fig.~\ref{flowchart}. For each Requesting BS-Requested BS pair the polling process entails the information exchange about resources and load status \cite{2013}. In the standard, the X$2$-AP defines two Elementary Procedures (EP) for this same purpose, namely the ``Resource Status Initiation" and ``Load Indication" procedures \cite{2013}. The former is defined as a class 1 EP (i.e., it consists of two messages, a request and a response, namely ``X$2$-AP:RESOURCES STATUS REQUEST" and ``X$2$-AP:RESOURCE STATUS RESPONSE" messages), whereas the latter is defined as a class 2 EP (i.e., it consists of a single message, without response, namely ``X$2$-AP:LOAD INFORMATION" message). RENEV makes use of these two EPs, defined by the X$2$-AP, to implement the detection phase.

As shown in Fig.~\ref{fig_metasignalling}(a), the Requesting BS sends the standardized ``X$2$-AP:RESOURCE STATUS REQUEST" message to the Requested BS (Fig.~\ref{fig_metasignalling}, message $1$) asking for the following information (known as IE in the X$2$-AP nomenclature): the percentage of RBs in use, the load on S$1$ interface and the hardware load. The Requested BS returns a response and then reports each IE for both uplink and downlink with the standardized ``X$2$-AP:RESOURCE STATUS RESPONSE" message (Fig.~\ref{fig_metasignalling}, message $2$) \cite{2013o}. Also, Load Indication procedure is used to transfer interference co-ordination information between neighboring BSs managing intra-frequency cells. The standardized ``X$2$-AP:LOAD INFORMATION" message (Fig.~\ref{fig_metasignalling}, message $3$) includes three IEs for the controlling cell: the transmitted power in every downlink RB, the interference received in every uplink RB, and the list of uplink RBs in which the BS intends to schedule distant mobiles \cite{2013o}. These control messages are necessary before transferring additional control information for establishing common RRC layer among BSs with RENEV. This procedure is repeated for all neighboring SCs. If none of the requested SCs has enough unused resources, the procedure is repeated with the eNB. Up to this point, all messages used by RENEV in the detection phase are defined in the standard \cite{2013o}.

\subsection{Transfer phase signaling}

We define the transfer of resources as the reconfiguration of a set of unused subcarriers to be vacated by the Donor BS and subsequently used by the Recipient BS. As this procedure is not considered in X$2$-AP, a new Class 1 EP compatible with the standard should be defined. In this paper the two proposed messages of the new EP are the messages $4$ and $5$ (Fig.~\ref{fig_metasignalling}). We denote them as ``X$2$-AP:METASIGNALLING INFORMATION REQUEST" and ``X$2$-AP:METASIGNALLING INFORMATION ACKNOWLEDGE", although other possible implementations are not precluded.

Once the Donor BS is selected, the initiating ``X$2$-AP:METASIGNALLING INFORMATION REQUEST" message (message $4$ in Fig.~\ref{fig_metasignalling}) is transmitted from Requesting BS to the Requested BS to show that resources are required by the former. The message must contain the following IEs: Message Type, Requesting BS X$2$-AP ID, Requested BS X$2$-AP ID and the corresponding transparent container. These IEs indicate the number of necessary RBs to cover the needs of the UE, and the identities of the Requesting and Requested BSs. For its part, the Donor BS returns a response to the Recipient BS via ``X$2$-AP:METASIGNALLING INFORMATION ACKNOWLEDGE" message (see Fig.~\ref{fig_metasignalling}, message $5$). This message carries all control information needed to execute the actual transfer of resources. The corresponding IEs are the Message Type, Cause, Bearers Admitted List, Bearers Rejected List and the equivalent transparent container. These IEs are necessary to confirm that the requested RBs exist in the Donor BS and that they are available for use by the Requesting BS.

\subsection{Discussion on the Time Scale of RENEV}

One dimension regarding the time scale of RENEV, is related to its duration. The algorithm is triggered every time that a Requesting BS lacks resources. The main RRC functions that have to be triggered per BS, RAC and RBC, are Layer 3 RRM functions. Therefore, the time scale of the algorithm resides on the time scale that RAC and RBC need in order to be activated in each BS. Another dimension of the time scale of RENEV, regards its periodicity of triggering. To begin with, it is expected that too often triggering of RENEV leads to excessive signalling of message exchange. In the second place, parameters such as the number of users or their mobility affect the signalling burden exchanged by RENEV. The ability of exchanging messages over X2 interface resides on the actual implementation of the interface (i.e., over the air wireless, fiber etc.). These are design parameters by the infrastructure owner. To conclude, although frequent triggering of RENEV leads to better adaptation to traffic variations, it also leads to higher message exchange over X2 interface, thereby increasing exponentially the signaling.

\section{Throughput Analysis} \label{Throughput Analysis}

\subsection{System Model} \label{System_Model}

As described in Section~\ref{IIIA}, the scenario consists of a single macro eNB (hereafter denoted as $BS_0$ and located at the center of the scenario) and a SCs tier, made up of a set of SC clusters, each one consisting of $N \in \mathbb{N}$ SCs (denoted as $BS_i$, with 1 $\leq i \leq N$), randomly distributed on a two-dimensional Euclidean plane $\mathbb{R}^2$. As clusters are not overlapped, there is no loss of generality in assuming one SC cluster within the eNB coverage area, creating a set of $N$+1 BSs, which is referred to as $B = \{ BS_i : 0 \leq i \leq N  \}$. We denote as $X \in \mathbb{N}$ the number of the overall users within the deployed scenario. These $X$ users are divided into $N$+1 traffic layers, each one geographically spanned over the coverage area of a BS. The coverage area of a $BS_i$ is defined as the region where users are served by this specific BS and all the users are assumed to be connected to the BS from which they receive the best Signal-to-Noise-Ratio (SNR), given that there are available RBs within $BS_i$. Given the described scenario, if the proportion of users contained within the coverage area of $BS_i$ is denoted as $a_i$, the number of users within this coverage area may be expressed as $X_i = a_i X$, with $\sum^{N}_{i=0} a_i =1$. Within each traffic layer, users are distributed uniformly.

\subsection{General Throughput Formulation} \label{throughput_analysis}

The LTE-A standard defines a discrete set of Modulation and Coding Schemes (MCSs) with the following possible configurations in the downlink for data transmission for both SCs and the eNB: QPSK ($\frac{1}{8}$, $\frac{1}{5}$, $\frac{1}{4}$, $\frac{1}{3}$, $\frac{1}{2}$, $\frac{2}{3}$, $\frac{3}{4}$), $16$-QAM ($\frac{1}{2}$, $\frac{2}{3}$, $\frac{3}{4}$) and $64$-QAM ($\frac{2}{3}$, $\frac{3}{4}$, $\frac{4}{5}$) \cite{2013d}. Based on a target bit error rate, the MCS is selected by the BS according to the $\text{SNR}$ received by the user. In that sense, given that the transmission rate depends on the applied MCS, the expected transmission rate per RB of a user connected to $BS_i$ is
\begin{equation} \label{eq1}
\mathbb{E}[R_i] = \sum_k P(\text{MCS}_i = k) \cdot R_{ik},
\end{equation}
\noindent where $P(\text{MCS}_i = k)$ is the probability of using the $k^{th}$ MCS in $BS_i$, and $R_{ik}$ is the transmission rate (in bps) achieved within a single RB with the $k^{th}$ MCS. The derivation for $P(\text{MCS}_i = k)$ may be found in Appendix~A. Note that \eqref{eq1} is valid for eNB and SCs. However, due to the overlapping of the coverage areas of the SCs and the eNB, the users located within the coverage area of a SC could be connected to the eNB if the available resources allocated to SCs do not suffice. In other words, a user of the $i^{th}$ traffic layer (with $i \neq 0$) could get connected to $BS_0$ despite $\text{SNR}_i > \text{SNR}_0$. Hence, if a user within the $i^{th}$ traffic layer (with $i \neq 0$) is served by the eNB, the expected transmission rate per RB is given by 
\begin{equation} \label{eq2}
\mathbb{E}[R_i^0] = \sum_k P(\text{MCS}_i^0 = k) \cdot R_{ik},
\end{equation}  
\noindent where $P(\text{MCS}_i^0 = k)$ is the probability of using the $k^{th}$ MCS in the eNB (i.e., $BS_0$) with a user in the $i^{th}$ coverage area ($i \neq 0$). Based on this, for a given number of users $X_i$, there is a group of users associated to $BS_i$, namely $X_i^i$, and a group of users associated to $BS_0$, denoted by $X_i^0$. Thus, for a given $X_i$, the expected number of users associated to $BS_i$ is

\begin{equation} \label{eq2a}
\mathbb{E}[X_i^i] = \min\bigg( X_i ,\frac{RB_i \cdot \mathbb{E}[R_i]}{d}  \bigg),
\end{equation} 
\noindent where $RB_i$ is the number of RBs allocated to $BS_i$ and $d$ is the specific demand of every single user (in bps). According to \eqref{eq2a}, $\mathbb{E}[X_i^i] = X_i$ if $RB_i$ is enough to serve all the attached users. Otherwise, not all users associated to $BS_i$ will be served. The maximum number of users that can be served is calculated from the expected maximum throughput, defined as the expected throughput per RB (i.e. $E[R_i]$) multiplied by the number of available RBs, $RB_i$. Thus, the expected maximum number of users is $\frac{RB_i \cdot E[R_i ]}{d}$. Finally, by definition, $\mathbb{E}[X_i^0] = X_i - \mathbb{E}[X_i^i] $. According to the definition, the total throughput, expressed as the sum of the throughput of each BS (i.e., $T = \sum_{i} T_i$), depends on the number of users from every operator connected to each BS, the transmission rate per RB, as well as the amount and the distribution of the available resources. In the following, we assume the use of a first-come first-served policy in each BS. This policy is equivalent to an extreme case of PRR, where $100$\% of the resources in each BS are shared and delivered on-demand (hereafter denoted as PRR $100$\%). This assumption (with and without RENEV) results in the upper bound of the aggregate throughput.

\subsection{Aggregate Throughput with RENEV}

Although RENEV negotiates resources in a peer-to-peer fashion among BSs, the procedure can be stochastically modelled as a single pool of resources dynamically allocated to the tenant BSs, when RENEV and PRR 100\% are implemented.Let us denote the throughput served by the eNB and generated by the $X_0$ users, as $T_{R,0}$. This throughput will equal the traffic generated by $X_0$ users associated to $BS_0$, subject to the availability of sufficient resources (i.e., $RB_0$). Thus,
\begin{equation}
T_{R,0} = \min\bigg( X_0 \cdot d, \mathbb{E}[R_0] \cdot RB_0 \bigg).
\label{eqtp1}
\end{equation}

Note that, when RENEV is applied, the eNB tends to transfer resources to the SCs, if necessary and feasible, rather than serve users within the coverage area of the SCs. Therefore, $X_i^0=0$ for $\forall i\neq0$. In turn, SCs serve their users with all the resources allocated within the SCs tier, as well as with unused resources in the eNB, $RB_0$. The application of RENEV may be modeled with two unified pools of resources; one composed of the RBs belonging to the SCs tier (denoted as $RB_T = \sum_{i \neq 0} RB_i$) and one consisting of the RBs from the eNB. Each Requesting BS will receive proportionally to its traffic load, resources from the SCs pool (i.e., $\frac{a_i}{1-a_0} \cdot RB_T$) and the corresponding portion of resources belonging to the eNB pool, denoted as $\mathbb{E}[RB_i^s]$. Therefore, the aggregate throughput generated by the SCs tier, according to the proof provided in Appendix~B, can be written as
\begin{equation}
\sum_{i\neq0} T_{R,i} = \min\bigg( X \cdot (1-a_0) \cdot d, \sum_{i \neq 0} \mathbb{E}[R_i] (\frac{a_i \cdot RB_T}{1-a_0}  + \mathbb{E}[RB_i^s] ) \bigg).
\label{eq_renev_th}
\end{equation}

\noindent Consequently, the expected overall system throughput with RENEV, is given by: $T_R = T_{R,0} + \sum_{i\neq0} T_{R,i}$.

\subsection{Aggregate Throughput without RENEV} \label{Ag_throuphut_without}

Alternatively, when RENEV is not applied (still considering a first-come first-served policy per BS, or in other words PRR 100\%), there is not any mechanism to reallocate resources, and consequently all BSs can only serve users with their initially allocated RBs. Similarly to \eqref{eqtp1}, the throughput of each SC is $T_{NR,i} = \min(X_i \cdot d, \mathbb{E}[R_i] \cdot RB_i), \text{ } \forall i \neq 0$.

As for the eNB throughput, it is divided into two components: the throughput offered by the $X_0$ users within the coverage area of $BS_0$ (i.e., $T_{NR,0}^0$); and the traffic offered by users within the coverage area of the SCs that cannot be served by these BSs due to lack of resources (i.e., $T_{NR,SCs}^0$ ):
\begin{equation}
T_{NR,0}^0 = \min\bigg(X_0 \cdot d, \mathbb{E}[R_0] \cdot RB_0 \bigg)
\label{eqnr00}
\end{equation}

\begin{equation}
T_{NR,SCs}^0 = \min\bigg(\sum_{i \ne 0} \mathbb{E}[X_i^0] \cdot d, \mathbb{E}[R_i^0] \cdot \left(RB_0 - \frac{T_{NR,0}^0}{\mathbb{E}[R_0]} \right) \bigg).
\label{eqnri0}
\end{equation}
\noindent According to \eqref{eqnri0}, if the available resources by the eNB (i.e., $RB_0$) are enough to serve the users in the coverage area of the SCs that cannot be served by them due to lack of RBs (i.e.,$\mathbb{E}[X_i^0]$ with $i \neq 0$), then they are served and their throughput equals $T^0_{NR,SCs} = \sum_{i \neq 0} \mathbb{E}[X_i^0] \cdot d$. Otherwise, not all $\mathbb{E}[X_i^0]$ users are served. The maximum throughput that can be achieved is calculated from the expected maximum throughput per RB (i.e., $\mathbb{E}[R_0^i]$), multiplied with the available remaining RBs in the SC tier. To calculate the latter, we subtract from the total number of RBs belonging to the eNB ($RB_0$), the ones used to serve the eNB's traffic (i.e., $\frac{T^0_{NR,0}}{\mathbb{E}[R_0]}$). Therefore, the aggregate throughput without RENEV is,
\begin{equation} \label{Ag_throuphut_without_eq}
T_{NR} = ( T_{NR,0}^0 + T_{NR,SCs}^0 ) + \sum_{i\neq0} T_{NR,i}.
\end{equation}

\section{Additional Signaling Overhead Analysis} \label{signaling analysis}

The densification of the network via the deployment of numerous SCs poses challenges in the infrastructure. Specifically, the need for a backhaul to interconnect BSs and forward both data traffic and signaling has emerged as one of the key points that could constrain the feasibility of these scenarios. Focusing on the implementation of RENEV, the whole communication among BSs relies on the existence and capacity of the logical X$2$ interface (as described in Section~\ref{Feasibility}). Although this logical interface is standardized \cite{2013}, the description of the backhaul physical infrastructure in order to support it, is left open. For such a reason, it is crucial from the infrastructure provider's perspective to assess the additional overhead introduced in the network by RENEV. In the following, we theoretically derive the number of signaling messages exchanged during RENEV operation, as well as the expression for the percentage of successful resources' transfer requests.

Given the system model presented in Section~\ref{System_Model} and the nomenclature used in Section~\ref{Nomenclature RENEV}, each BS may be characterized by the number of RBs initially allocated to it as well as the number of used/unused RBs for a particular number of users. Thus, let us define, the number of available resources for a specific $BS_i$ as $r_i = RB_i-u_i$, where $RB_i$ are the RBs initially allocated to $BS_i$ and $u_i$ is the number of RBs required to serve the demand of the users associated to $BS_i$. The number of required resources, $u_i$, will be upper and lower bounded as a function of the number of users connected to $BS_i$, their traffic demand and their received SNR. Therefore, $u_i \in [ u_{i,min} , u_{i,max} ]$, where $u_{i,min}$ and $u_{i,max}$ are the numbers of RBs being required when all the UEs associated to $BS_i$ use $64$QAM$\frac{4}{5}$ (i.e., the maximum throughput per RB) and QPSK$\frac{1}{8}$ (i.e., the minimum throughput per RB) respectively. Based on these definitions, the upper and lower bounds of available resources for the set of BSs of the described system can be defined as $r_{min} = \underset{0 \leq i \leq N}{\min} ( RB_i - u_{i,max} )$ and $r_{max} = \underset{0 \leq i \leq N}{\max} ( RB_i - u_{i,min} )$.

In this context, the system is defined by the set of possible initial states $\mathcal{S} = \{ S_1 , S_2, \ldots , S_W \}$ and the set of probabilities of occurrence of each state $\pi = \{  \pi_1 , \pi_2 , \ldots , \pi_W \}$, where $W$ stands for the number of possible states. In turn, each state is defined as $S_j = ( s_{j,1}, s_{j,2}, \ldots ,  s_{j,r_{max}-r_{min}+1} )$, where $s_{j,k} \in \mathbb{N}_0$ denotes the sum of BSs with a number of available resources equal to $(r_{min}-1+k)$ and $S_j \in \mathcal{S}$. As in RENEV the BSs first seek for resources in the SCs tier and subsequently in the eNB, we decouple the analysis into these two steps. Focusing first on the SCs tier (without considering the resources in the eNB), the system may be defined by the set of possible initial states $\mathcal{S}$ and the probability of occurrence $\pi$. By definition, $\sum_{k=1}^{r_{max}-r_{min}+1} s_{j,k} = N$. According to the definitions stated above, the number of Requesting BSs in a given state $S_j$, will be equal to the number of BSs with negative $r_i$, also expressed as $n_R(S_j) = \sum_{k=1}^{-r_{min}} s_{jk}$.
 Therefore, the expected number of Requesting BSs may be written as
\begin{align} \label{eq_req_BSs}
\mathbb{E} [n_R] = \sum_{j=1}^{W} n_R (S_j ) \cdot \pi_j.
\end{align}
After the operation of RENEV in the SCs tier, the available resources of the Donor BSs will have been transferred to the Requesting BSs to cover their needs. Consequently, the probability of having the system in a particular state $S_j$ after executing RENEV will vary. If we denote by $\pi_j'$ the probability of being in the state $S_j$ after the RENEV completion in the SCs tier, it holds,
\begin{equation} \label{prob_aft_ren_sc}
\pi_j' = \sum_{n=1}^{W} \pi_n \cdot p_{nj},
\end{equation}
\noindent where $p_{nj}$ is the probability of transiting from state $S_n$ to $S_j$. Note that not all transitions are feasible since the redistribution of resources among SCs imposes some restrictions. Thus, $p_{nj}\neq0$ if and only if $S_j$ is contained in the set of feasible future states of $S_n$, i.e., $S_j \in \mathcal{F}(S_n)$. The detailed definition of $\mathcal{F}(S_n)$, according to the conditions that should hold to satisfy that $S_j \in \mathcal{F}(S_n)$, is introduced in Appendix~C. Hence, the transition probability, is given by
\begin{align} \label{transition_prob_s1}
p_{nj}=\left\{
    \begin{array}{ll}
	   1 							&: j = n, \mathcal{F}(S_n) = \emptyset \text{,}\\ 
       \frac{1}{|\mathcal{F}(S_n)|} &: j \neq n, S_j \in  \mathcal{F}(S_n) \text{,}\\ 
       0 							&: \text{otherwise,} 
     \end{array}
   \right.
\end{align}
\noindent where $|\mathcal{F}(S_n)|$ is the cardinality of the set $\mathcal{F}(S_n)$. Although the SCs tier is the first alternative for RENEV to reallocate the existing resources, not all requests can be covered with the resources of this tier. Thus, and according to \eqref{eq_req_BSs}, the expected number of successful requests (i.e., when the needs of the Requesting BSs are covered by the unused resources of the Donor BSs) in the SCs tier may be calculated as
\begin{align} \label{suc_requests}
\mathbb{E} [n_s] = \sum_{j=1}^{W} n_R (S_j ) \cdot [  \pi_j - \pi_j'].
\end{align}


As RENEV is completed in the SCs tier, all feasible redistribution of resources has been successfully conducted, and the system is found in state $S_j \in \mathcal{S}$, with probabilities $\pi'$. However, note that $S_j$ characterizes the scenario without taking into account the resources available in the eNB, i.e., $r_0$. Therefore, in the second step of the signaling analysis a new set of states, namely $\mathcal{S}''$, must be defined to include $r_0$. It should be noted that the $r_0$ resources inserted into the system, may be distributed in different ways. For instance, if all Requesting BSs are overlapped among them, the new resources will be transferred to the SCs tier only once. Conversely, if not all Requesting BSs overlap with the rest of the Requesting BSs, the $r_0$ resources will be transferred more than once. Therefore, if we define the number of non-overlapping groups of Requesting BSs as $Q= \{ 1, 2 \cdots M \}$, where $M$ stands for the number of Requesting BSs (for instance, for $S_j$ we have $M=n_R(S_j)$, the $r_0$ resources can be transferred to the SCs tier $Q$ times. Thus, for a specific state $S_j$ containing $M$ Requesting BSs, the inclusion of the $r_0$ resources from the eNB can lead to $M$ possible new states. Specifically, a state $S_j$ results in $M$ new states defined as $S''_t = (s''_{t,1}, s''_{t,2}, \ldots , s''_{t,k}, \ldots,  s''_{t,r_{max}-r_{min}+1} )$, with $s''_{t,k}=s_{j,k}+Q$ for $k = r_0 - r_{min} + 1$ and $Q= \{ 1, 2 \cdots M \}$, and $s''_{t,k}=s_{j,k}$ otherwise. This set of new states is defined for each value of $r_0$. Therefore, after the inclusion of the resources available in the eNB the system may be described by the set of new possible initial states $\mathcal{S}^{''}= \{S''_1 , S''_2 , \ldots , S''_L \}$ and the probability of being initially in these states $\pi'' = \{\pi^{''}_1 , \pi^{''}_2 , \ldots , \pi^{''}_L \}$, where $L$ stands for the number of possible states. Thus, it holds that

\begin{equation}
\pi^{''}_t = \pi_j' \cdot P(Q=q|N,M) \cdot P_{eNB}(r_0), \label{prob_pi_double_prima}
\end{equation}
\noindent where $P(Q=q|N,M)$ is the probability of having $q$ non-overlapping groups in a cluster with $M$ Requesting BSs out of $N$ BSs (calculated in Appendix~D) and $P_{eNB}(r_0)$ is the probability that the eNB has $r_0$ spare RBs that could be transferred. For a given scenario, the latter is a random variable that depends on the resources allocated to the eNB, the number of users and the traffic demand of each user.  
 
Henceforth, we use the same calculation method that we used for the SCs tier to derive the expected number of successful requests. Firstly, the expected number of Requesting BSs is calculated as in \eqref{eq_req_BSs}, using the new probabilities of occurrence $\pi^{''}$, denoted as $\mathbb{E}[n'_R] =\sum_{j=1}^L n_R (S''_j ) \cdot \pi''_j$. After the application of RENEV, the available resources of the eNB will have been transferred to the Requesting BSs. The new transition probabilities from state $S''_n$ to $S''_j$ for this phase, according to \eqref{prob_aft_ren_sc}, will be equal to $\pi^{'''}_j = \sum_{n=1}^{L} \pi_n'' \cdot p'_{nj}$, \noindent where $p_{nj}'$ is calculated with \eqref{transition_prob_s1} and the set of feasible future states $\mathcal{F}(S_n'')$ according to Appendix~C. Under the conditions stated above, it cannot be assured that all requests can be covered with the resources of the eNB tier. Thus, the expected number of successful requests in the eNB tier may be calculated as $\mathbb{E} [n'_s] = \sum_{j=1}^{L} n_R (S''_j ) \cdot [  \pi^{''}_j - \pi^{'''}_j]$. Therefore the total expected number of successful requests by both tiers after the completion of RENEV is equal to $\mathbb{E}[n_{s_{total}}] = \mathbb{E} [n_s] + \mathbb{E} [n'_s]$, and the probability of successful requests is calculated as $( \frac{\mathbb{E}[n_{s_{total}}]}{\mathbb{E}[n_R]})$. The number of signaling messages exchanged by the BSs depends on the total number of BSs (i.e., $N+1$), the number of Requesting BSs, and the number of successful requests. In particular, and by observing Fig.~\ref{fig_metasignalling}, it can be noticed that all Requesting BSs (whose number is in average equal to $\mathbb{E}[n_R]$) exchange $3$ messages (messages $1$, $2$ and $3$) with the rest of the $N-1$ SCs. Additionally, the Requesting BSs not being able to obtain resources from the SCs tier (whose number is in average $ \mathbb{E}[n'_R] $) exchange the aforementioned three messages with the eNB. Finally, if any of the requests is successful, the Requesting BSs exchange $2$ messages (messages $4$ and $5$ in Fig.~\ref{fig_metasignalling}). Therefore, the expected number of signaling messages exchanged by RENEV may be expressed as: $\mathcal{I} = 3 \cdot (N-1) \cdot \mathbb{E}[n_R] + 3\cdot \mathbb{E}[n'_R] + 2\cdot \mathbb{E}[n_{s_{total}}]$.

\section{Performance Evaluation} \label{Performance Evaluation}

\subsection{Simulation Scenario and Parameters}

The number of clusters per eNB coverage area can vary from $1$ to optional $4$ and the number of SCs per cluster can vary from $1$ to $10$ depending on the actual deployment \cite{2013g}. Therefore, our simulation scenario consists of an eNB overlaid with a cluster of SCs, consisting of $6$ outdoor HeNBs-LTE femtocells \cite{6516167} \cite{Small_Cell_forum}, operating on the same carrier frequency \cite{2013f,2013g}. We conducted Monte-Carlo extensive simulations (with a thousand iterations to achieve statistical validity) in a custom made simulation tool implemented in MATLAB\textsuperscript{\textregistered}, using random deployments of a SCs cluster placed within the eNB coverage area. In each iteration mobile users are distributed independently and non uniformly; i.e., $2/3$ are dropped within the SC tier \cite{2013f,2013g}. The simulation parameters are listed in Table~\ref{parameters}; the $3$GPP related parameter values are based on \cite{2013i}. The overall system bandwidth consists of $2$ bands of $20$ MHz, operating at $2$ GHz, each one assigned to each tier using CA. Packet scheduling is proportional fair both at eNB and SCs. We conduct simulations for a full buffer traffic model\cite{2013g}. Users download files using File Transfer Protocol (FTP) at an average data rate of $300$ Kbps.

\begin{table}[htbp]
\renewcommand{\arraystretch}{0.99}
\caption{Basic System Parameters used in the Simulation}
\label{parameters}

\centering
\begin{tabular}{|c | c|}
\hline
\textbf{Parameters} & \textbf{Settings/Assumptions}\\
\hline
Network layout & Cluster of $6$ HeNB LTE Femtocells \\
			   & randomly placed per Macrocell\\
\hline
Inter-site    & Macrocell: $500$ m (ISD)\\
distance/cell radius 	   & Femtocell: $25$ m (Cell radius)\\
\hline
Transmit power & Macrocell: $46$ dBm,
			     Femtocell: $17$ dBm \\
\hline
Bandwidth	   & $20$ MHz at $2$ GHz for each tier\\
\hline
Path loss &	Macrocell: $140.7+36.7log10$(R[km])\\
	      & Femtocell: $128.1+37.6log10$(R[km])\\
\hline
Shadow fading &	Lognormal, $\mu$ = 0, std.=$8$ dB for Macrocell\\
			  & Lognormal, $\mu$ = 0, std.=$10$ dB for Femtocell\\
\hline
\end{tabular}

\end{table}

As discussed in the previous sections, RENEV is a complementary virtualization solution implementable on top of existing solutions. Hence, in the scenario under consideration both NVS \cite{Costa-Perez2013} and PRR \cite{Guo2013} are simulated with and without RENEV. NVS creates distinct slices of spectrum in each particular BS. These slices accommodate equal percentage of the overall RBs, each one residing in a specific traffic flow. PRR framework, guarantees a minimum number of RBs per subframe on average for each traffic flow, which is available when a particular flow wants to use it (i.e., reserved part). The portion of system resources remaining after subtracting the reserved part at each BS, is called shared part and it can be used by any incoming traffic flow. According to \cite{Guo2013}, an operator requires at least a minimum portion of resources to be reserved for its users within a BS, in order to guarantee QoS for particular traffic slices. In simulations, for users downloading FTP files this percentage is set to $50$\% \cite{Guo2013} corresponding to the scheme named PRR $50$\%. Although setting a high value for shared part within a BS can lead to more flexible allocation of resources, it comes with the shortcoming of not covering the minimum requirements for QoS imposed by operators. However, we use this maximum degree of flexibility in PRR, having $0$\% RBs reserved part and $100$\% shared within each BS (i.e., ``PRR $100$\%"), to calculate the theoretical upper bound of the aggregate throughput.

\subsection{Network Performance} \label{Network Point of View}

\begin{figure}[htbp]
	\centering
	\includegraphics[width=0.8\textwidth]{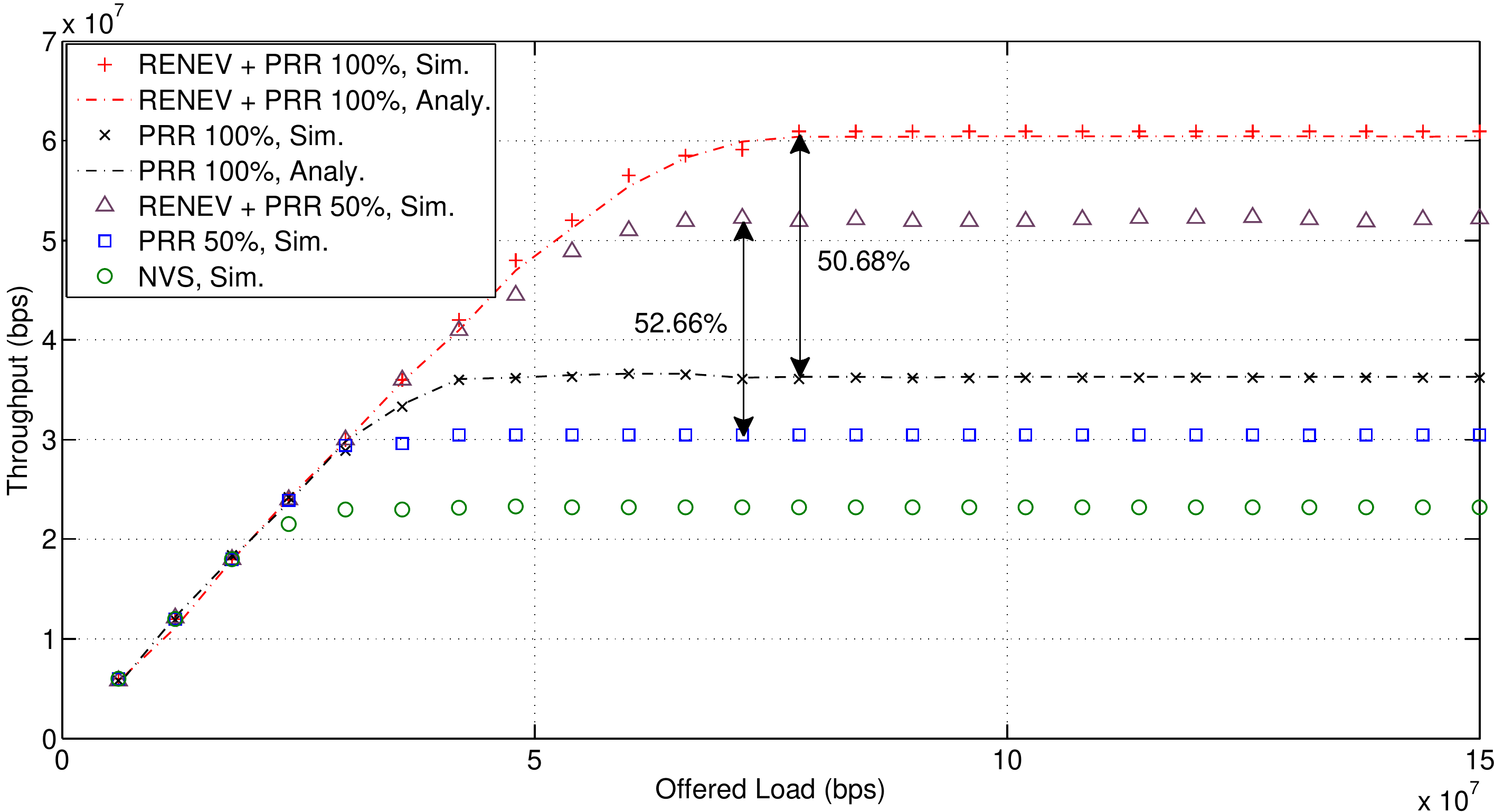}
	\caption{Aggregate System Throughput for different number of Offered Loads.}
	\label{fig_comparison}
\end{figure}

Fig.~\ref{fig_comparison} presents the aggregate system throughput (a metric indicated by $3$GPP in \cite{2013f,2013i}) with respect to an increasing offered traffic load for NVS as well as PRR $50$\% and PRR $100$\% with and without RENEV. As it may be observed, the experimental and theoretical curves for PRR $100$\% and RENEV+PRR $100$\% (the upper bound expressions as derived in Section~\ref{Throughput Analysis}) match. For offered load equal to $18$ Mbps, the system's behavior is the same for all the depicted schemes; all demanded traffic is served. However, as the load increases all compared schemes are able to serve less users compared to the system where RENEV is applied. In particular, when saturation is reached due to a lack of resources (i.e., offered load equals $78$ Mbps), the throughput achieved with RENEV + PRR $100$\% ($60.93$ Mbps) represents an increase of $50.68$\% with respect to PRR $100$\%. In the first case, the available resources of two tiers are distributed according to traffic demand to cover the maximum number of users' needs; however when RENEV is not applied, each BS manages its own resources which are depleted after a while. At the other extreme, the NVS scheme achieves the poorest performance, since resources from different slices cannot be shared regardless of the traffic demands in each slice. The maximum value in this case is $23.19$ Mbps. As for PRR $50$\%, with and without RENEV, its performance constitutes an intermediate situation. 

Notwithstanding the good results offered by PRR $100$\% compared to PRR $50$\% (both of them without the application of RENEV), the authors in \cite{Guo2013} expound that a minimum share of the available resources should be reserved for each traffic slice to guarantee minimum QoS requirements. Therefore, PRR $100$\% is not convenient in terms QoS despite outperforming PRR $50$\% in terms of aggregate throughput. The same conclusion applies when RENEV is implemented. By inspecting Fig.~\ref{fig_comparison}, it is particularly worth noting that RENEV + PRR $50$\% (which does not degrade the QoS requirements of the traffic slices) is able to show higher aggregate throughput than PRR $100$\%. This behavior is due to the ability of RENEV to compensate not only the traffic spatial non-uniformities but also the QoS loss experienced when sharing the $50$\% of the resources per BS, instead of the $100$\% in PRR.

\begin{figure}[htbp]
	\centering
	\includegraphics[width=0.8\textwidth]{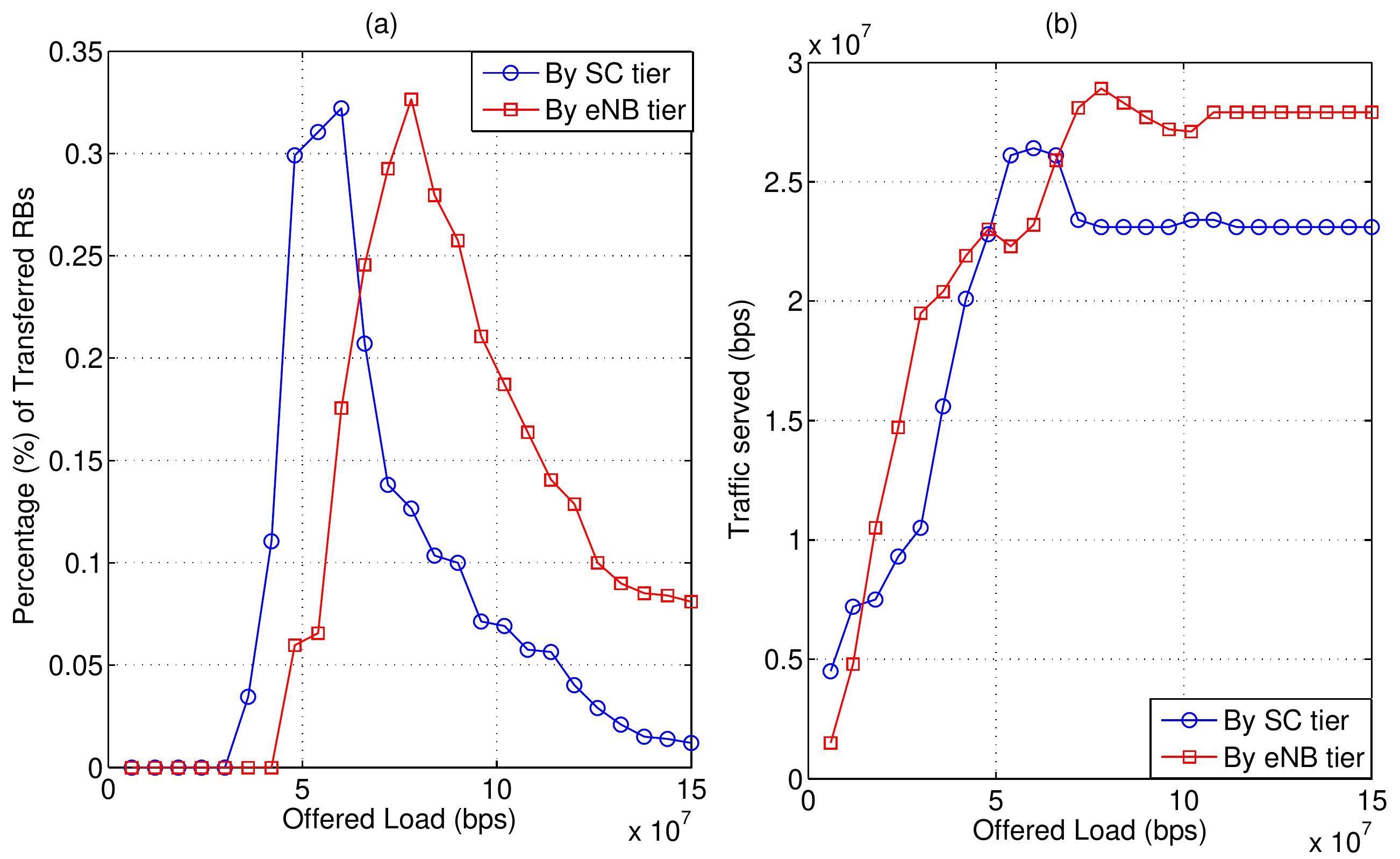}
	\caption{(a) Percentage of transferred RBs by each tier. (b) Traffic Served by each tier. }
	\label{fig_percentage}
\end{figure}

Figures \ref{fig_percentage}(a) and \ref{fig_percentage}(b) study the percentage of transferred RBs per tier as well as the corresponding served traffic for the case of RENEV + PRR $100$\% (as depicted in Fig.~\ref{fig_comparison}). As expected, we observe that the RB transfer first increases, then reaches a specific peak and then decreases for both tiers. The two peaks in Fig.~\ref{fig_percentage}(a) equal $32.2$\% of transferred RBs by the SCs tier (achieved for $60$ Mbps) and $32.64$\% by the eNB (achieved for $78$ Mbps). After these peaks, although the number of users requiring resources is augmented, the transferred resources decrease because both tiers run out of RBs since all of them are already allocated to the existing users. It is worth noting that the traffic served by each tier (Fig.~\ref{fig_percentage}(b)) depends on the available number of RBs. In particular, when the percentage of transferred RBs falls, the aggregate throughput in Fig.~\ref{fig_comparison} stabilizes since the resources are depleted and the incoming user requests cannot be satisfied. 

Finally, when applying RENEV, the resources are provided to the tenant Requesting BSs first by the SCs tier and subsequently, when none of the SCs is able to provide resources, by the eNB, that acts as a donor BS. For this reason, we may observe fluctuation points for the served traffic, among $50$ Mbps and $100$ Mbps in Fig.~\ref{fig_percentage}(b). In particular, for low traffic load, most transfer of resources is conducted among the SCs. Progressively, as offered traffic increases, it is less probable that SCs provide additional resources. Thus the eNB starts transferring resources to the Requesting BSs. When the maximum load is achieved in the SCs tier (i.e., $60$ Mbps), the probability of finding a Donor BS within this tier falls. On the same time, the eNB (which is still less loaded than the SCs) keeps increasing the percentage of transferred resources till $78$ Mbps. At this point, the eNB is also loaded and the probability of transferring to Requesting BSs decreases. This is translated into the served traffic; the traffic served by the SCs grows thanks to the transfer of resources from the SCs tier and from the eNB. However, when the transfer of resources by the SC tier falls, the increase of eNB transfer of RBs cannot compensate it and the traffic served by the SCs tier decreases. Due to high load in the eNB tier as well, when the transfer of resources decreases, the traffic tends to stabilize to the maximum traffic that can be served by the SCs tier without the transfer of resources. The served traffic by the eNB is also stabilized to the maximum value that can be served by it.

\subsection{User's Throughput} \label{User Point of View}

\begin{figure}[htbp]
	\centering
	\includegraphics[width=0.8\textwidth]{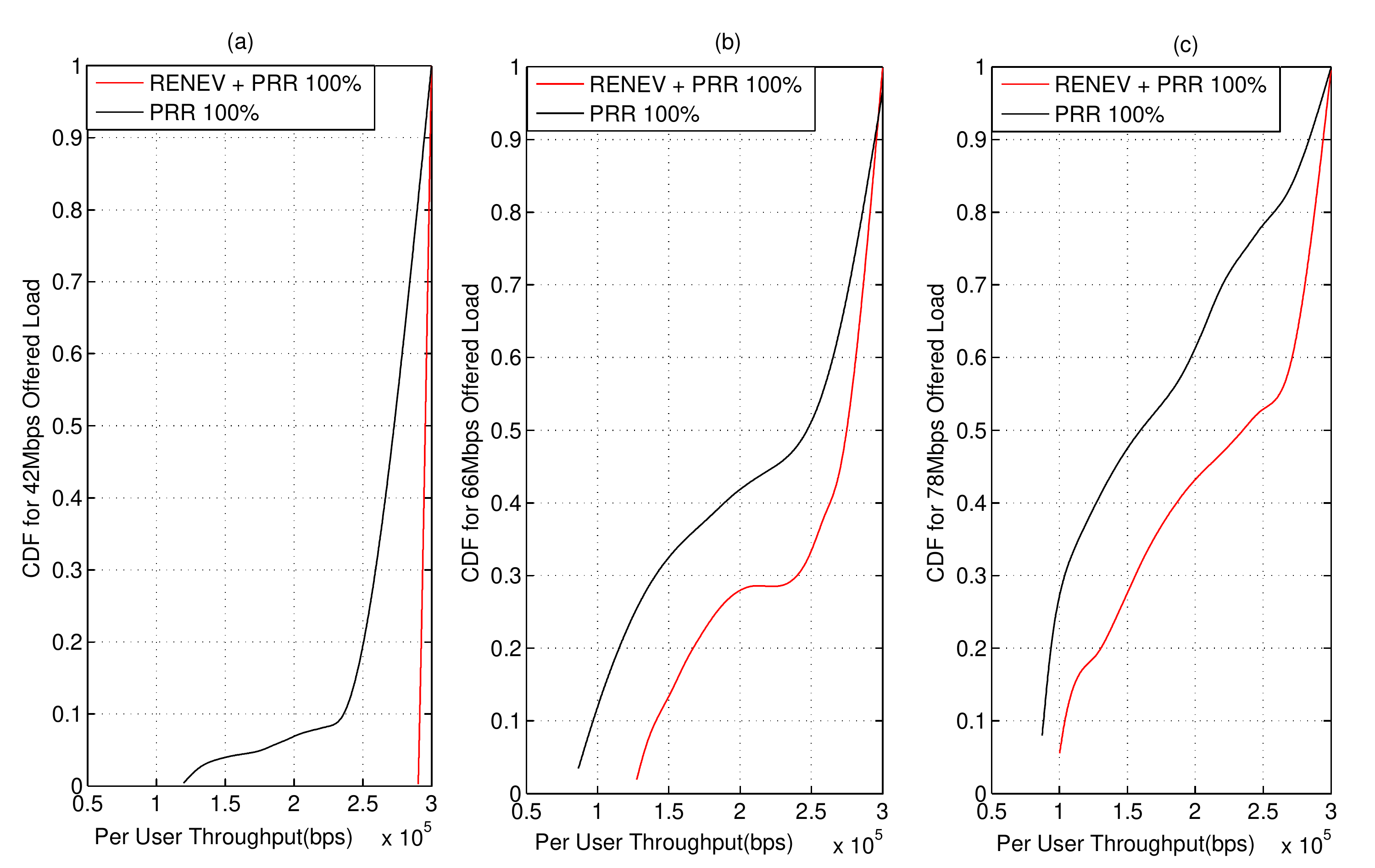}
	\caption{CDF of user Throughput for (a) 42Mbps, (b) 66Mbps and (c) 78Mbps Offered Load.}
	\label{fig_CDF_42}
\end{figure}

In Figures \ref{fig_CDF_42}(a), \ref{fig_CDF_42}(b) and \ref{fig_CDF_42}(c) we study the Cumulative Distribution Function (CDF) of user throughput (indicated metric in \cite{2013f,2013i}) for three cases of traffic load: low offered load where the majority of users are served, medium one and the case where the system is saturated; $42$ Mbps, $66$ Mbps and $78$ Mbps correspondingly, as also depicted in Fig.~\ref{fig_comparison}. In the sequel focus on the scenario with PRR $100$\% with and without RENEV, since it provides the upper bounds of network's throughput. First, we observe that the gains in throughput acquired in the network side with the application of RENEV, can be translated into merits for the end users. According to Fig.~\ref{fig_CDF_42}(a), as the offered load is low, RENEV is able to help the majority of users to achieve the demanded data rate. In particular, the observed slight deviation from $300$ Kbps, is due to the fact that some users do not achieve the demanded data rate because of the channel conditions that they experience. However, without applying RENEV the user throughput dispersion is quite high. For instance, $80$\% of the users achieve throughput values equal or higher to $250$ Kbps. The rest $20$\% of the users achieve values ranging from $120$ Kbps to $250$ Kbps.

In addition, we observe that higher offered load affects dramatically the user throughput. For example, in Fig.~\ref{fig_CDF_42}(b), $72$\% of the users achieve transmission rate equal or higher than $250$ Kbps when RENEV is applied. On the other hand for the same percentage without applying RENEV the lowest user throughput achieved is $130$ Kbps. In particular, the transfer of resources defined by RENEV, improves the performance of users with poor links, who are normally located in the cell edge area. These users are more demanding in terms of required RBs. However, RENEV is able to satisfy such kind of users. For instance, when the system is further loaded (Fig.~\ref{fig_CDF_42}(c)) the dispersion among user throughput is quite high, both with and without RENEV. Even in this study case, $50$\% of the overall users achieve $75$\% of the demanded transmission rate (with lowest user throughput equal to $102$ Kbps). On the contrary, without RENEV, this percentage falls to $52.5$\% of the demanded data rate.

\subsection{Signaling Overhead}

In this set of our experiments, we evaluate the requests and the corresponding messages that are necessary for the transition from a scenario where all resources are initially distributed uniformly among the BSs, to a scenario where the resources are finally distributed according to the existing geographical traffic variations (i.e., upper bound values).


\begin{figure*}[htbp]
\hfill
\subfigure[]{\includegraphics[width=0.49\textwidth]{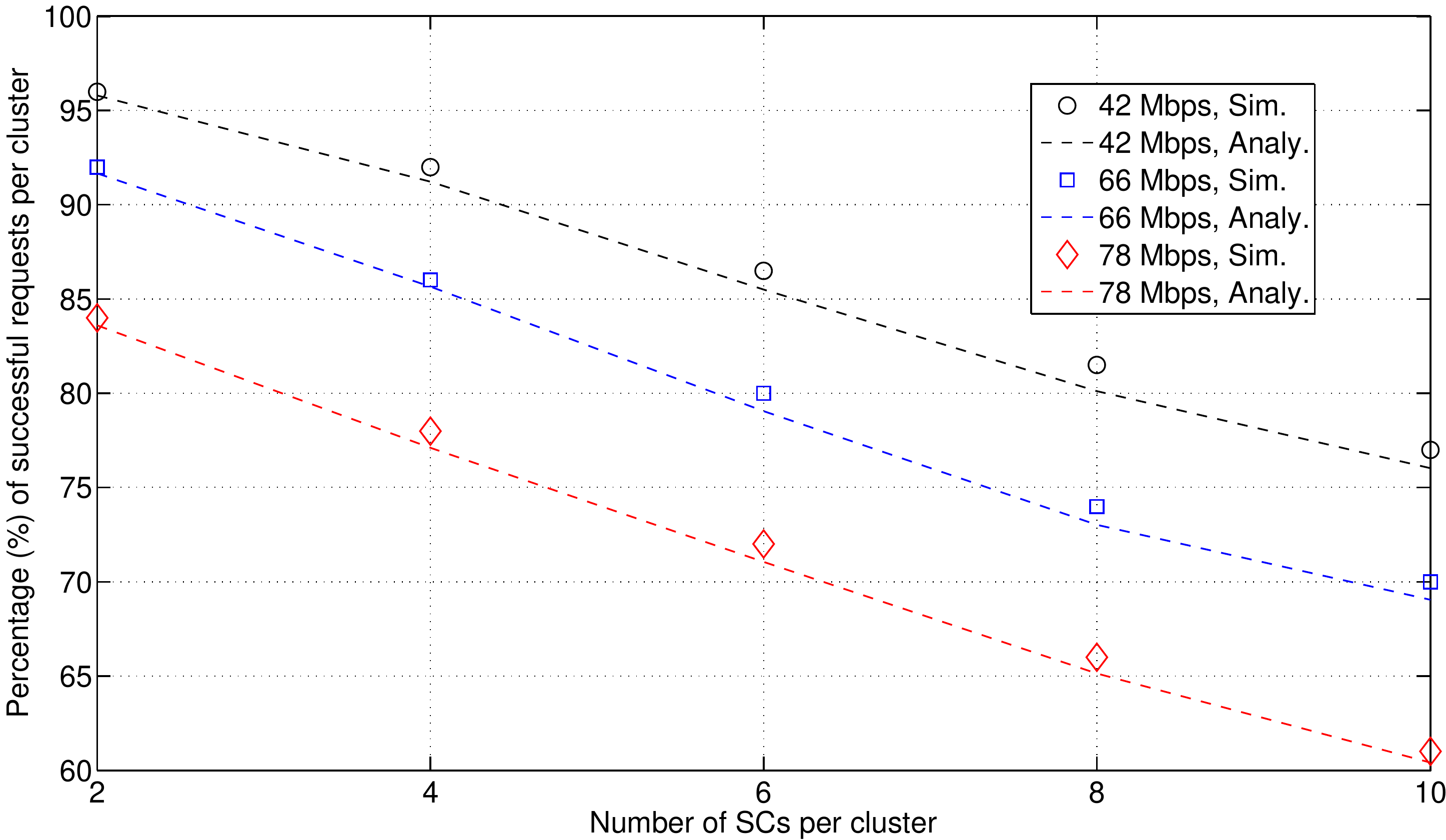}}
\hfill
\subfigure[]{\includegraphics[width=0.46\textwidth]{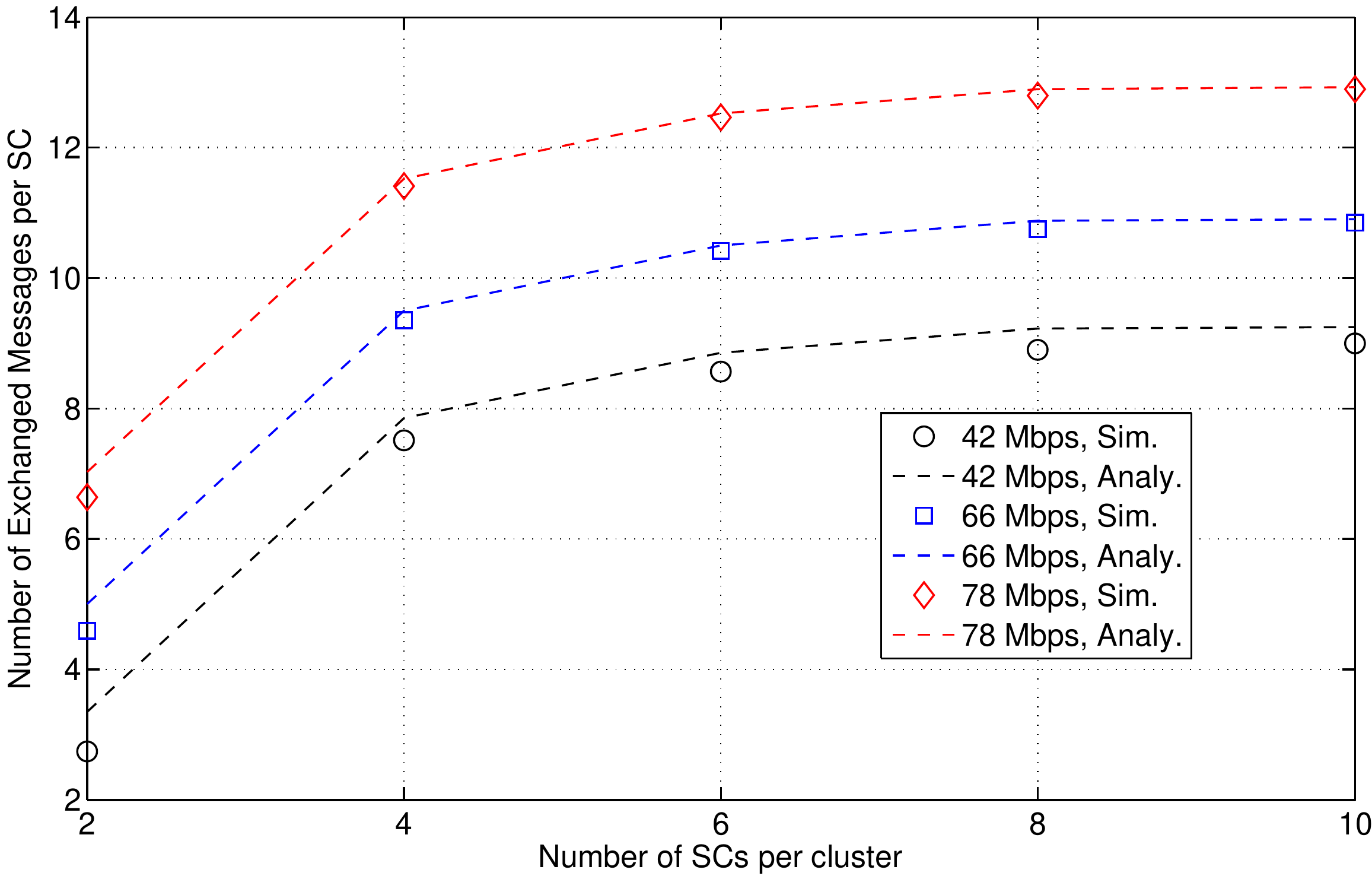}}
\hfill
\caption{(a) Percentage of successful requests for different number of SCs per cluster. (b) Number of exchanged X$2$ messages per SC.}
\label{signalling}
\end{figure*}

In Fig.~\ref{signalling}(a) we study the impact of the number of SCs into the percentage of successful requests per cluster, for different traffic offered loads (low, medium, and high as in Fig.~\ref{fig_CDF_42}). It is worth noting that in dense scenarios in terms of SCs, the available RBs are quickly depleted, and therefore, the number of successful requests falls. This means that the tenant Requesting BSs cannot attain the demanded resources. For high loaded systems less requests are satisfied since resources are exhausted faster. For example, if a cluster with $6$ SCs is considered (scenario analyzed in Fig.~\ref{fig_comparison}), the percentage of successful requests is $86.5$\% for $42$ Mbps offered load, $80$\% for $66$ Mbps and $72$\% for $78$\% Mbps. On the other hand, when $10$ SCs are considered within the cluster's surface, this percentage falls to $77$\%, $70$\% and  $61$\%, respectively.

Fig.~\ref{signalling}(b) studies the number of exchanged messages per SC, for the three studied offered loads. In all cases, the experimental results showcase that higher number of SCs within the cluster, is translated into higher number of exchanged messages over X$2$ interface. For instance, for a cluster with $6$ SCs, we observe in average $8.5$ exchanged messages for $42$ Mbps, $10.4$ for $66$ Mbps and $12.4$ for $78$ Mbps. In particular, as the number of SCs in a cluster increases, the messages among the participant tenant BSs are also increasing even though the rate of increase progressively reduces.  

The physical implementation of X$2$ is still not standardized, so it should be noted that it is the main factor imposing feasibility constraints. In general we note that a particular number of SCs where RENEV can be applied depends on the limits inserted of the actual implementation of X$2$ and the corresponding capacity reserved for signaling. Fig.~\ref{signalling}(b) can result quite useful for operators, to calculate the actual signaling for a certain number of SCs per cluster, according to the way they choose to implement X$2$ (i.e., such as fiber, over-the-air wireless, etc.).

\section{Concluding Remarks} \label{Conclusion}

In this paper, we have proposed RENEV; a scheme that considers the coordination among several BSs to create an abstraction of systems' radio resources, so that multiple tenants (i.e., BSs) can be served, in a heterogeneous environment. The extensive performance assessment has revealed that gains in system's throughput are translated into gains for the users' throughput as well. With the use of RENEV, system's resources are dynamically distributed according to users' needs on an isolated and on-demand basis. In this way, the majority of the users is served, as long as spare resources exist. Finally, the solution has been evaluated for the signaling overhead that adds into the network for increasing number of SCs per cluster.

\appendices

\section{MCS Selection Probability} \label{app_mcs}

Let us denote by $x_{i} \in \mathbb{R}^2$ the location of $BS_i$ and $y \in \mathbb{R}^2$ a random location in the scenario. The signal strength received from $BS_i$ at location $y$, expressed in dB, may be written as $p_{i}(y) = P_{T_i} - L_i (y) - S_i ( y)$, where $P_{T_i}$ is the constant that includes antenna gains and transmitted power of $BS_i$, $L_i (y)$ is the path loss from $x_i$ to $y$, and $S_i (y)$ is the slow fading. The $\text{SNR}$ received at $y$ from $BS_i$, when no interference is received, is given by $\text{SNR}_i(y)_{dB} = p_i(y) - N_0$, where $N_0$ represents the noise average power. Throughout the rest of the analysis, taking into account the transmission power and coverage area of each BS as well as that subcarriers are not utilized by neighboring cells, we assume that interference is imperceptible among them \cite{2013f}. Without loss of generality, the dependency of the several variables on the location $y$ will be omitted in the sequel. Yet, all expressions are still derived for a random location $y$. Therefore, let $\text{SNR}_{max}$ be the highest SNR received from a BS in $B$ at a random location $y$, where $\text{SNR}_{max} = \max_{BS_i \in B} \text{SNR}_i$.


Focusing on the adaptive MCS mechanism, the $k^{th}$ MCS is selected by $BS_i$ if and only if $\text{SNR}_k^{min} \leq \text{SNR}_i < \text{SNR}_k^{max}$, where $\text{SNR}_k^{min}$ and $\text{SNR}_k^{max}$ stand for the minimum and maximum thresholds of MCS $k$, respectively. Therefore, the probability of using a certain MCS could be expressed as:

\begin{align} 
P(\text{MCS}_i = k)  
			&= \frac{P(\text{SNR}_k^{min} \leqslant \text{SNR}_i < \text{SNR}_k^{max} \cap \text{SNR}_i=\text{SNR}_{max})}{P(\text{SNR}_i=\text{SNR}_{max})}.
\label{eqa}			
\end{align}

Since the SNR of a particular BS$_i$ is considered independent from the SNR of the rest BSs, 

\begin{align}
P(\text{SNR}_i = \prod_{j \ne i} P(\text{SNR}_i>\text{SNR}_j) =\prod_{j \ne i} P(S_j>S_i + \mu_{ij}),
\label{eqb}
\end{align}
\noindent where $\mu_{ij}=P_{T_j}-P_{T_i}+L_i-L_j$. Based on the analysis provided in \cite{Tseliou2014(tobepresented)} and after a convenient change of variables, \eqref{eqb} is equal to $ F_{S_i}(\frac{\sigma_i  \cdot \mu_{ij}}{\sigma_j  \cdot \sqrt{2}})$, where $F_{S_i}$ denotes the Cumulative Distribution Function (CDF) of the random variable $S_i$ expressing the shadowing, whereas $\sigma_i $ and $\sigma_j $ denote the standard deviations of the shadowing of $BS_i$ and $BS_j$. Correspondingly, the numerator of \eqref{eqa} is derived in \cite{Tseliou2014(tobepresented)} as $P(\text{SNR}_k^{min} \leqslant \text{SNR}_i < \text{SNR}_k^{max} \cap \text{SNR}_i=\text{SNR}_{max}) = \prod_{j \neq i} P(\text{SNR}_k^{min} \leqslant \text{SNR}_i < \text{SNR}_k^{max} \cap \text{SNR}_i > \text{SNR}_j)$. By substituting the values $S^0 = P_{T_i} - SNR_k^{max} - L_i$ and $S^1 = P_{T_i} - SNR_n^{min} - L_i$, the previous equation is expressed as follows:

\begin{align}
P(S^0 \leqslant S_i < S^1 \cap S_j > S_i + \mu_{ij}) = (F_{s_i}(S^1) - F_{s_i}(S^0)) - \int_{S^0}^{S^1} F_{s_i} (s_i + \mu_{ij})f_{s_i}(s_i)ds_i.
\end{align}
\noindent where $ f_{s_i}(s_i)$ is the Probability Distribution Function (PDF) of $S_i$.

\section{Derivation of Equation (6)} \label{eq6}

For deriving the throughput achieved by the users located within the SCs tier with RENEV, let us divide the process according to the source that provides RBs to the Requesting BSs. First resources are redistributed within the SCs tier to serve the demanded traffic. In the case that these are not enough, resources are granted from the eNB. To begin with, SCs tier redistributes its RBs to accommodate the demanded traffic. If the overall traffic is less or equal to the SCs capacity, all users can be served. The overall resources within this tier, are equal to $RB_T = \sum_{i \neq 0} RB_i$. What is more, the average transmission rate for this case, equals $\mathbb{E}[R_{TOT}] = \frac{1}{1-a_0} \cdot \sum_{i \neq 0} a_i \cdot \mathbb{E}[R_i]$, where $a_i$ denotes the percentage of users located within the coverage area of $BS_i$ and $\mathbb{E}[R_i]$ the expected transmission rate in $BS_i$. Thus, if $\sum_{i \neq 0} X_i \cdot d \leq RB_T \cdot \mathbb{E}[R_{TOT}]$, all users located in the SC tier (i.e., $\sum_{i \neq 0}X_i = X \cdot (1- a_0)$) will be served by the SCs' resources. It follows that $\sum_{i \neq 0} T_{R_i} = d \cdot \sum_{i \neq 0} X_i$.


Once SCs' resources (i.e., $RB_T$) are depleted, $X \cdot (1-a_0)$ users within the Requesting BSs, will require further resources from the eNB tier. Therefore, the expected number of users to be served with resources from the eNB is $\mathcal{E} = X \cdot (1-a_0) - \frac{RB_T}{d} \cdot \mathbb{E}[R_{TOT}]$. Thus, each Requesting $BS_i$ will have to serve $\mathcal{E}_i = \frac{a_i}{1-a_0} \cdot \mathcal{E}$ users. Let us denote as $\mathbb{E}[RB_{i}^{s}]$, the amount of resources from the eNB that can be given to each Requesting $BS_i$.

Based on this, a particular Requesting $BS_i$, will serve all this traffic (i.e., $\mathcal{E}_i \cdot d$) in the case where $\mathcal{E}_i \cdot d \leq \mathbb{E}[RB_{i}^{s}] \cdot \mathbb{E}[R_i]$ holds. In contrast, the traffic served by Requesting $BS_i$ with RBs from the eNB tier will be equal to $\mathbb{E}[RB_{i}^{s}]  \cdot \mathbb{E}[R_i]$. If $\mathcal{E}_i \cdot d$ is served, in total the throughput achieved within this tier, will equal $\sum_{i \neq 0}\mathcal{E}_i \cdot d =  d \cdot \mathcal{E}$. Otherwise, it will be yielded by the summation of the traffic served in each Requesting BS with resources from the eNB (i.e., $\sum_{i \neq 0} \mathbb{E}[RB_{i}^{s}]  \cdot \mathbb{E}[R_i]$). Consequently the throughput generated by the users in the SCs tier, served both with resources redistributed within the SCs tier and resources transferred from the eNB, will be equal to

\begin{equation} \label{eq32}
\sum_{i \neq 0} T_{R_i} = \min \bigg ( X \cdot (1 -a_0) \cdot d, RB_T \cdot \mathbb{E}[R_{TOT}] + \sum_{i \neq 0} \mathbb{E}[RB_i^s] \cdot \mathbb{E}[R_i] \bigg ).
\end{equation}
It remains to show how we calculate the number of eNB resources, that can be given to each Requesting BSs (i.e., $\mathbb{E}[RB_{i}^{s}]$), included in \eqref{eq32}. For the sake of simplicity and without loss of generality, we can assume circular cluster's surface containing $N$ circular shaped SCs. Therefore, the area of the cluster is $A = \pi \cdot R_c^2$, where $R_c$ is the cluster radius, and $A_i = \pi \cdot R_{SC_i}^2$ holds for a circular shaped coverage area with Radius $R_{SC_i}$. For a particular Requesting $BS_i$, located randomly within the cluster, there will be an overlap if the distance between $BS_i$ and another $BS_j$ is less than $R_{SC_i}+R_{SC_j}$. Thus the probability of overlap among two Requesting BSs is derived as

\begin{equation} \label{poverlap}
P_o = \frac{\pi \cdot (R_{SC_i}+R_{SC_j})^2}{\pi \cdot R_c^2} = \bigg( \frac{R_{SC_i}+R_{SC_j}}{R_c} \bigg) ^2.
\end{equation}
Then, the probability for a Requesting $BS_i$ of having $n_i$ overlaps is described by a binomial random variable as follows:

\begin{equation} \label{ni_overlaps}
P(n_i = n) = \binom{N-1}{n} \cdot P_o^n \cdot (1-P_o)^{N-1-n}.
\end{equation}
\noindent Although log-normal shadowing is considered, our assumption of circular coverage SCs has been validated by simulations (for $17$ dBm  SC transmission power, $\mu=0$ dB and $\sigma_S=10$ dB as indicated in Table~\ref{parameters}). We assume that a Requesting $BS_i$ with $n_i$ overlapping BSs, receives$( RB_s \cdot \frac{1}{n_i + 1} )$ RBs. The expected value of this term is equal to

\begin{align} \label{exp_value_first}
\mathbb{E}[\frac{RB_s}{n_i+1}] &= \sum_{n_i=0}^{N-1} \frac{RB_s}{n_i+1} \cdot \binom{N-1}{n_i} \cdot P_o^{n_i} \cdot (1-P_o)^{N-1-n_i}, 
\end{align}
where a convenient change of variables can be applied, $m=n_i+1$, so as \eqref{exp_value_first} equals

\begin{align}  \label{RBs_SCs_same}
\mathbb{E}[\frac{RB_s}{n_i+1}] &= \frac{RB_s}{P_o} \cdot \sum_{m=1}^{N} \frac{(N-1)!}{m!(N-m)!} \cdot P_o^m \cdot (1-P_o)^{N-m}=\frac{RB_s}{NP_o} \cdot [1 - (1-P_o)^N],
\end{align}
\noindent which is valid for all Requesting BSs since each one is assumed to receive $\frac{RB_s}{n_i+1}$. However this is not true in the case that each Requesting BS accommodates different portion of users (i.e., $a_i$). This difference in the Requesting BS load, implies different traffic demands and hence unequal percentage of resources to be allocated. Let us assume, as previously, that Requesting $BS_i$ overlaps with $n_i$ BSs. The number of the possible ways of overlapping equals $\binom{N-1}{n_i}$. Each of these ways can occur with probability $(P_o^{n_i} \cdot (1-P_o)^{N-1-n_i})$. Let us define the set $\mathcal{O}_{ic}^{n_i}$ as a particular set of $n_i$ overlapping Requesting BSs with Requesting $BS_i$. All Requesting BSs in $\mathcal{O}_{ic}^{n_i}$, as well as Requesting $BS_i$, will share RBs according to the proportion of users that each one accommodates. Thus, the percentage of resources achieved per Requesting $BS_i$ is equal to $\frac{a_i}{a_i + \sum_{BS_k \in O_{ic}^{n_i}}a_k}$. Therefore
\begin{align} \label{RBs_SCs_diff}
\mathbb{E}[RB_{i}^{s}] &= RB_s \cdot \sum_{n_i=0}^{N-1} P_o^{n_i} \cdot (1-P_o)^{N-1-n_i} \cdot \sum_{c=1}^{\binom{N-1}{n_i}} \frac{a_i}{a_i + \sum_{BS_k \in \mathcal{O}_{ic}^{n_i}}} 
\end{align}

\section{Set of Feasible Future States} \label{appendix}

RENEV is intended to redistribute the unused resources of the possible Donor BSs among the Requesting BSs. Therefore, not all transitions from state $S_n$ to state $S_j$ are feasible. The set of feasible future  states for a given state $S_n$, $\mathcal{F}(S_n)$, is defined as the set of states to which $S_n$ could transit after performing RENEV. Based on the definition of states $S_n$, $S_j$ and RENEV algorithm, the following conditions must be accomplished to assure that $S_j \in \mathcal{F}(S_n)$:
\begin{itemize}
\item The amount of resources is constant in the initial and the final states: $\sum_{k=1}^{r_{max}-r_{min}+1} s_{j,k} \cdot (r_{min}-1+k) = \sum_{k=1}^{r_{max}-r_{min}+1} s_{n,k} \cdot (r_{min}-1+k)$.
\item After performing RENEV, the number of Requesting BSs should be smaller. Therefore, $n_{R}(S_j) < n_{R}(S_n)$.
\item The number of requested RBs in the final state should be less than the corresponding number in the initial state: $\sum_{k=1}^{-r_{min}} s_{jk} \cdot  (r_{min}-1+k) < \sum_{k=1}^{-r_{min}} s_{nk} \cdot (r_{min}-1+k)$.
\item In state $S_j$ (i.e., final state) there are not new Requesting BSs. Therefore, $\forall s_{n,k} = 0$ and $k \leq -r_{min}$, then $s_{j,k}=0$. Likewise, $\forall s_{n,k} \neq 0$ and $k \leq -r_{min}$, it holds that $s_{j,k}\leq s_{n,k}$.  

\item The number of RBs transferred by the Donor BSs is equal to the number of RBs received by the Requesting BSs: $\sum_{k=2-r_{min}}^{r_{max}-r_{min}+1} (s_{n,k}-s_{j,k}) \cdot (r_{min}-1+k) = \sum_{k=1}^{-r_{min}} (s_{n,k}-s_{j,k}) \cdot (-r_{min}+1-k)$.
\item The absolute value of the highest amount of requested RBs in the initial state (i.e., negative value) should be lower or equal than the minimum amount of available RBs such that if $k' = r_{max}-r_{min}+1 \text{ , }\forall k \leq -r_{min} \text{ , } \forall s_{n,k} \neq 0 \text{ then } s_{jk} \neq 0 \text{ , } \forall k \geq |r_{min}-1+k'|$.
\item As RENEV is completed, all possible redistribution of resources has been done. Therefore, there is not any possible Donor BS that could cover the needs of a Requesting BS. Thus, $\forall s_{j,k} \neq 0$ and $k \leq -r_{min}$, it is true that $\sum_{m=-2r_{min}+2-k}^{r_{max}-r_{min}+1} s_{j,m} =0$.

\end{itemize}

\section{Derivation of $P(Q=q | N,M)$} \label{app_overlaps}

The probability that two BSs within the cluster are overlapping is derived in \eqref{poverlap}, denoted as $P_o$, and the probability that a specific $BS_i$ in the cluster is overlapped with $n$ BSs (no overlapping among different clusters is assumed), denoted as $P(n_i=n)$, is derived in \eqref{ni_overlaps}. Note that, for a given state $S_j$, if $BS_i$ is assumed to be a Requesting BS, the probability that a BS different from $BS_i$ is a Requesting BS equals $P_{N,M}=\frac{M-1}{N-1}$, where $M=n_R (S_j)$. Let us denote with $m_i$, the number of Requesting BSs overlapping $BS_i$. Henceforth, $P_{RB}(m_i = m |N,M)$ denotes the probability that $m$ Requesting BSs overlap $BS_i$, given that $M$ out of $N$ BSs are Requesting BSs it can be expressed as

\begin{equation}
P_{RB}(m_i = m | N,M) =  \sum_{k=m}^{N-1} \bigg ( P(n_i=k) \cdot \binom{k}{m} \cdot P_{N,M}^m \cdot ( 1 - P_{N,M} ) ^{k-m} \bigg ).
\end{equation}
In RENEV, the eNB will only transfer the same resources to two different Requesting BSs if they do not overlap. Approximately, we could claim that the available resources of the eNB can be transferred to a specific SCs cluster, as many times as the number of non-overlapping groups of Requesting BSs. Therefore, we are interested in figuring out the number of non-overlapping groups of Requesting BSs within the cluster, denoted as $Q = \{1,2, \cdots M \}$. For instance, when all Requesting BSs overlap altogether, $Q=1$; when there are two non-overlapping groups of BSs, $Q=2$ (i.e., BSs are overlapped within each group but non-overlapped with the BSs of the other group); finally, when all Requesting BSs are not overlapped, $Q=M$. If we assume that all BSs within each group overlap with each other, the probability of having $Q$ non-overlapping groups of BSs can be approximated by

\small{
\begin{equation}
P(Q=q | N,M) \simeq \left\{
\begin{array}{ll}
P_{RB}(m_i=N-1|N,M) & \text{if } q=1 \text{,}\\
\sum_{k=0}^{M - 1} P_{RB}(m_i=k|N,M) \cdot  P_{RB}(m_i=M-q-k|N,M)) & \text{if } q=2 \text{,} \\
\sum_{k=0}^{M-Q} P_{RB}(m=k|N,M) \cdot P(Q=q-1|N-1-k,M-1-k) & \text{if } q > 2 \text{.} 
\end{array}
\right.
\end{equation}}


\section*{Acknowledgment}
This work has been funded by the MITN Project CROSSFIRE (PITN-GA-2012-317126), by the AGAUR (2014SGR 1551) and by the CellFive (TEC2014-60130-P) Research Projects.




%
\begin{IEEEbiography}{Georgia Tseliou}
(S'13) received the M.Sc. degree in computer engineering from Computer Engineering and Informatics Department, University of Patras, Grecee in 2012. 

From 2010 to 2012, she was with the Computer Technology Institute and Press ``Diophantus" (CTI), Patras, Greece. She is currently a Marie Curie Early Stage Researcher at the Internet Interdisciplinary Institute (IN3) of Open University of Catalonia (UOC), pursuing a PhD diploma at the Signal Theory and Communications Department (TSC) of Universitat Politècnica de Catalunya BarcelonaTech (UPC). Her research has been concerned with Network Virtualization Techniques and Radio Resource Management in LTE-A environments.
\end{IEEEbiography}

\begin{IEEEbiography}{Ferran Adelantado}
(M'07) received the telecommunications engineering and PhD degrees from Polytechnic  University of Catalonia, Barcelona, in 2001 and 2007, respectively, and the BS degree in business sciences from Open University of Catalonia, Barcelona, in 2012. 

Since December 2007, he has been an Associate Professor with Open University of Catalonia, where he has participated in several European and National funded projects. His research interests include different aspects of wireless networks, particularly medium access control protocols and radio resources management. Currently, his research is mainly focused on LTE-A networks.
\end{IEEEbiography}


\begin{IEEEbiography}{Christos Verikoukis}
received the Ph.D. degree from UPC in 2000. He is currently a Senior Researcher at CTTC and an Adjunct Professor at the University of Barcelona. His area of expertise is in the design of energy-efficient layer 2 protocols and RRM algorithms for short-range wireless cooperative and network coded communications. He has published 82 journal papers and over 160 conference papers. He is also a co-author of three books, 14 chapters in other books, and two patents. He has participated in more than 30 competitive projects, and has served as the principal investigator of national projects in Greece and Spain as well as the technical manager in 5 projects. He has supervised 15 Ph.D. students and five postdoctoral researchers since 2004. He was General Chair of the 17th, 18th, and 19th IEEE CAMAD, the TPC Chair of the 6th IEEE Latincom, and TPC Co-Chair of the 15th Healthcom. He has also served as the CQRM symposium co-chair in the IEEE Globecom 2014 and IEEE ICC 2015 \& 2016. He is currently the Chair of the IEEE ComSoc Technical Committee on Communication Systems Integration and Modeling. He received the Best Paper Award of the Communication QoS, Reliability \& Modeling Symposium in IEEE ICC 2011 and in Selected Areas in Communications Symposium of the IEEE Globecom 2014 as well as the EURASIP 2013 Best Paper Award for the Journal on Advances in Signal Processing
\end{IEEEbiography}
\vfill

%
%
%




\end{document}